%% file: preprint.tex
\documentclass[numberedappendix]{emulateapj}
\usepackage{epsf}

\slugcomment{Version 2, submitted to ApJ}

\newcommand\vect[1]{{\mbox{\boldmath $#1$}}}
\newcommand\bby  {{\bf y}}
\newcommand\bme  {\vect{\eta}}
\newcommand\bmo  {{\bf 0}}
\newcommand\bt   {\vect{\theta}}
\newcommand\bu   {\vect{u}}
\newcommand\bx   {{\bf x}}
\newcommand\ep   {\varepsilon}
\newcommand\grad {\mathop{\rm grad}\nolimits}
\newcommand\jac  {\mathop{\rm Jac}\nolimits}
\newcommand\ha   {\hat{a}}
\newcommand\hb   {\hat{b}}
\newcommand\hc   {\hat{c}}
\newcommand\hp   {\hat{\rm p}}
\newcommand\hq   {\hat{\rm q}}
\newcommand\Rcusp{R_{\rm cusp}}
\newcommand\Rein {R_{\rm ein}}
\newcommand\ttp  {{\rm p}}
\newcommand\ttq  {{\rm q}}

\newcommand\farcs{\hbox{$.\!\!^{\prime\prime}$}}
\newcommand\arcsec{\hbox{$^{\prime\prime}$}}
\newcommand\arcdeg{\hbox{$^\circ$}}

\newcommand\eq[1]{eq.~(\ref{eq:#1})}
\newcommand\refsec[1]{\S\ref{sec:#1}}
\newcommand\refapp[1]{Appendix~\ref{app:#1}}
\newcommand\reffig[1]{Figure~\ref{fig:#1}}
\newcommand\reffigs[2]{Figures~\ref{fig:#1} and \ref{fig:#2}}
\newcommand\reftab[1]{Table~\ref{tab:#1}}

\begin{document}

\title{Identifying Lenses with Small-Scale Structure. I. Cusp Lenses}

\author{
  Charles R.~Keeton\altaffilmark{1,2},
  B.~Scott Gaudi\altaffilmark{1,3},
  and A.~O.~Petters\altaffilmark{4,5}
}

\altaffiltext{1}{Hubble Fellow}
\altaffiltext{2}{
  Astronomy and Astrophysics Department, University of Chicago,
  Chicago, IL 60637; {\tt ckeeton@oddjob.uchicago.edu}
}
\altaffiltext{3}{
  School of Natural Sciences, Institute for Advanced Study,
  Princeton, NJ 08540; {\tt gaudi@sns.ias.edu}
}
\altaffiltext{4}{
  Department of Mathematics, Duke University, Durham, NC 27708;
  {\tt petters@math.duke.edu}
}
\altaffiltext{5}{Department of Physics, MIT, Cambridge, MA 02139}

\begin{abstract}

The inability of standard models to explain the flux ratios in
many 4-image gravitational lens systems has been presented as
evidence for significant small-scale structure in lens galaxies.
That claim has generally relied on detailed lens modeling, so it
is both model dependent and somewhat difficult to interpret.  We
present a more robust and generic method for identifying lenses
with small-scale structure.  For a close triplet of images created
when the source lies near an ideal cusp catastrophe, the sum of
the signed magnifications should exactly vanish, independent of
any global properties of the lens potential.  For realistic cusps,
the magnification sum vanishes only approximately, but we show
that it is possible to place strong upper bounds on the degree to
which the magnification sum can deviate from zero.  Lenses with
flux ratio ``anomalies,'' or fluxes that significantly violate
the upper bounds, can be said with high confidence to have
structure in the lens potential on scales of the image separation
or smaller.  Five observed lenses have such flux ratio anomalies:
B2045+265 has a strong anomaly at both radio and optical/near-IR
wavelengths;
B0712+472 has a strong anomaly at optical/near-IR wavelengths
and a marginal anomaly at radio wavelengths;
1RXS~J1131$-$1231 has a strong anomaly at optical wavelengths;
RX~J0911+0551 appears to have an anomaly at optical/near-IR
wavelengths, although the conclusion in this particular lens
is subject to uncertainties in the typical strength of octopole
density perturbations in early-type galaxies;
and finally, SDSS~J0924+0219 has a strong anomaly at optical
wavelengths.
Interestingly, analysis of the cusp relation does not reveal a
significant anomaly in B1422+231, even though this lens is known
to be anomalous from detailed modeling.  Methods that are more
sophisticated (and less generic) than the cusp relation may
therefore be necessary to uncover flux ratio anomalies in some
systems.  Although these flux ratio anomalies might represent
either milli-lensing or micro-lensing, we cannot identify the
cause of the anomalies using only broad-band flux ratios in
individual lenses.  Rather, the conclusion we can draw is that
the lenses have significant structure in the lens potential on
scales comparable to or smaller than the separation between the
images.  Additional arguments must be invoked to specify the
nature of this small-scale structure.

\end{abstract}

\keywords{cosmology: theory --- dark matter --- galaxies: formation
--- gravitational lensing --- large-scale structure of universe}

\section{Introduction}

Gravitational lens modeling has had remarkable success handling
increasingly precise measurements (e.g., Barkana et al.\ 1999;
Patnaik et al.\ 1999; Trotter, Winn, \& Hewitt 2000) and increasingly
sophisticated datasets including Einstein ring images (Keeton et
al.\ 2000; Kochanek, Keeton, \& McLeod 2001) and/or stellar dynamical
data (Romanowsky \& Kochanek 1999; Koopmans \& Treu 2002, 2003;
Treu \& Koopmans 2002, 2003; Koopmans et al.\ 2003b).  Lens modeling
has even clarified the properties of complex systems with more than
one lens galaxy and/or more than one background source (Cohn et al.\
2001; Rusin et al.\ 2001; Keeton \& Winn 2003; Koopmans et al.\ 2003b).
However, the notable success has largely been restricted to the number
and configuration of lensed images.  The flux ratios between the
images, at least in lenses with four or more images,\footnote{The
problem is less apparent in 2-image lenses, mainly because the limited
number of constraints leaves more freedom in the models.} have long
resisted explanation.

Until recently the persistent problem with flux ratios in 4-image
lenses (e.g., Kent \& Falco 1988; Falco, Leh\'ar, \& Shapiro 1997;
Keeton, Kochanek, \& Seljak 1997) received little attention, perhaps
because the number of 4-image lenses was relatively small, and because
it seemed possible to appeal to electromagnetic --- non-gravitational
--- effects such as extinction by dust or scattering by hot gas.
However, the number of lenses with apparently anomalous flux ratios
is growing rapidly (e.g., Inada et al.\ 2003; Sluse et al.\ 2003;
Wisotzki et al.\ 2003).  Moreover, direct evidence suggests that
electromagnetic effects, while present in some lenses, cannot explain
most of the anomalies (Falco et al.\ 1999; Winn et al.\ 2001, 2002;
Koopmans et al.\ 2003a).  The problem with flux ratios therefore
appears to be real.

It also turns out to have interesting and important implications for
astrophysics and cosmology.  When Mao \& Schneider (1998) made the
first systematic analysis of the flux ratio problem, they realized
that the anomalies might be attributed to gravitational effects
omitted from standard lens models, namely small-scale structure in
the lens galaxy.  The key insight was that since flux ratios are
determined by second derivatives of the lens potential, they are
much more sensitive to small-scale structure than the image positions
(which are determined by first derivatives of the potential); so
models that lack small-scale structure might successfully reproduce
the image positions but fail to fit the flux ratios.

One possible source of small-scale structure is clumps of dark
matter of mass $\sim\!10^6$--$10^9\,M_\odot$ left over from the
hierarchical galaxy formation process in the Cold Dark Matter (CDM)
paradigm (Metcalf \& Madau 2001; Chiba 2002; Dalal \& Kochanek 2002).
This possibility has generated significant interest because it relates
to current questions about the validity of CDM on small scales.  The
discrepancy between the predicted abundance of dark matter clumps and
the observed abundance of dwarf galaxy satellites around the Milky
Way has been interpreted as a fundamental problem with CDM (Klypin
et al.\ 1999; Moore et al.\ 1999), which may signal a need for new
physics for the dark matter (e.g., Spergel \& Steinhardt 2000; Colin,
Avila-Reese, \& Valenzuela 2000; Hu, Barkana, \& Gruzinov 2000).
Alternately, the discrepancy may simply indicate poor understanding
of the astrophysical processes that determine whether or not a clump
of dark matter hosts a visible dwarf galaxy (Bullock, Kravtsov, \&
Weinberg 2000; Benson et al.\ 2002; Somerville 2002; Stoehr et al.\
2002; Hayashi et al.\ 2002).  If lens flux ratios can be used to
probe dark matter clumps, that will provide the cleanest way to
distinguish these two very different hypotheses, and more generally
to resolve the controversy about whether CDM does or does not
over-predict small-scale structure (e.g., Flores \& Primack 1994;
Moore 1994; Spergel \& Steinhardt 2000; Debattista \& Sellwood 2000;
de Blok, McGaugh, \& Rubin 2001; Keeton 2001a; van den Bosch \&
Swaters 2001; Weiner, Sellwood, \& Williams 2001; de Blok \& Bosma
2002; Kochanek 2003).  Early results indicate that the statistics
of flux ratio anomalies imply a clump population that agrees well
with CDM predictions and validates cold dark matter (Dalal \&
Kochanek 2002; Kochanek \& Dalal 2003), but the importance of the
conclusion demands further study.

A second interesting possibility is that the small-scale structure
implied by flux ratio anomalies is simply stars in the lens galaxy
(Chang \& Refsdal 1979; Irwin et al.\ 1989; Wo\'zniak et al.\ 2000;
Schechter \& Wambsganss 2002).  In this case, flux ratio anomalies
offer a unique probe of the relative contributions of stars and dark
matter to the surface mass density at the image positions (Schechter
\& Wambsganss 2002), which would be interesting because the amount
of dark matter contained in the inner regions of elliptical galaxies
is still not well known (e.g., Gerhard et al.\ 2001; Keeton 2001a;
Borriello, Salucci, \& Danese 2003; Rusin, Kochanek, \& Keeton 2003b).
Yet a third possibility is that the small-scale structure is not
localized like dark matter clumps or stars, but is more global like
small disk components in bulge-dominated systems, Fourier mode
density fluctuations, tidal streams, etc.\ (e.g., Mao \& Schneider
1998; Evans \& Witt 2002; Quadri, M\"oller, \& Natarajan 2003;
M\"oller, Hewett, \& Blain 2003).  If this is the case, then lensing
can be used to search for such structures whether they are traced
by the luminous components of galaxies or not.

These three disparate applications all rest on a common foundation:
the identification of lenses with flux ratio anomalies that indicate
small-scale structure.  That identification is most unambiguous when
time variability (e.g., Wo\'zniak et al.\ 2000; Schechter et al.\
2003) or resolved spectra of the images (e.g., Moustakas \& Metcalf
2003; Wisotzki et al.\ 2003) clearly indicate microlensing by stars
in the lens galaxy, or when the resolved shapes of the images
indicate structure on the scale of dark matter clumps (e.g., Metcalf
2002).  Until such data become available for the majority of lenses,
however, we need a method to identify anomalies using only broad-band
flux ratios.  Besides, such a method will be needed to select
candidates for the expensive follow-up observations (monitoring,
spectroscopy, or high-resolution imaging).

To date, the usual approach has been to use detailed lens modeling
to interpret broad-band flux ratios and draw conclusions about,
for example, the abundance of dark matter clumps (e.g., Dalal \&
Kochanek 2002; Metcalf \& Zhao 2002; Kochanek \& Dalal 2003).
This approach is vulnerable to the criticism that the results
depend on the sorts of lens potentials used in the modeling.  The
argument has two parts.  First, many of the commonly used families
of lens potentials implicitly possess global symmetries, which
lead to invariant magnification relations that are ``global'' in
the sense that they involve all four images (Dalal 1998; Witt \&
Mao 2000; Dalal \& Rabin 2001; Hunter \& Evans 2001; Evans \& Hunter
2002).  If a fit is poor because the data fail to satisfy these
relations, that does not automatically constitute a flux ratio
anomaly; it may simply indicate that the assumed relations are
too restrictive, and that small, unremarkable deviations from the
assumed symmetries are needed.  The conceptual difficulty here is
that one is trying to use global relations to draw conclusions
about structure on smaller, more local scales.  The second part
of the argument is that there is a large difference in scale
between the image separations ($\sim\!0\farcs2$--$2\arcsec$) and
the scales relevant for dark matter clumps ($\sim\!10^{-3}\arcsec$)
or stars ($\sim\!10^{-6}\arcsec$).  If the flux ratio anomalies
are in fact due to structures that are intermediate between these
scales, then they may not necessarily imply the presence of dark
matter clumps or stars (Evans \& Witt 2002; Quadri et al.\ 2003;
M\"oller et al.\ 2003).

To address the first part of the criticism, we seek a method of
identifying flux ratio anomalies that is local rather than global,
i.e., a method that is sensitive only to structures smaller than
the scales probed by the image positions.  Fortunately, one can do
this by appealing to simple, generic relations between the image
magnifications that should be satisfied for images in ``fold'' or
``cusp'' configurations (defined in \refsec{configs}).  The
magnification relations are derived from local properties of the
lens mapping and are in principle independent of the global mass
model.  They can be violated only if there is significant structure
in the lens potential in scales smaller than the separations between
the images (see Mao \& Schneider 1998).  In practice, however, the
situation is complicated by the fact that the caustics in real lens
systems only approximate ideal folds and cusps in some low-order
expansion of the potential near the critical point; higher-order
terms introduce deviations from the fold and cusp geometries.
Real lenses therefore need to obey the ideal magnification
relations only approximately.  Because the accuracy with which the
relations should hold depends on the distance of the images from
the critical point and on properties of the lens potential, it is
not straightforward to judge {\it a priori\/} the significance of
an apparent violation.

Our goal is to understand the magnification relations in realistic
lens potentials and to determine how well they can be used to
identify flux ratio anomalies.  In this paper we focus on cusp
configurations, because as the highest order stable singularities
in lensing maps (see Schneider, Ehlers, \& Falco 1992; Petters,
Levine, \& Wambsganss 2001) cusps are amenable to analytic study,
and cusp configurations are easy to identify.\footnote{A close
triplet of images always indicates a cusp configuration; but a
close pair of images could be associated with either a fold or a
cusp.}  We will address fold configurations in subsequent work.
We study the degree to which the ideal cusp relation can be
violated due to various properties of the lens potential:  the
radial density profile, ellipticity, and multipole density
perturbations of the lens galaxy, and the external tidal shear
from the lens environment.  Using both analytic and numerical
methods we derive upper bounds on the deviation from the ideal
cusp relation for realistic lens potentials that lack significant
small-scale structure.  We then argue that finding larger
deviations in observed lenses robustly reveals flux ratio
anomalies and indicates the presence of some sort of small-scale
structure.

We assert that, even though we adopt specific families of lens
potentials, our analysis is more general than explicit modeling.
One reason is that we have a better distinction between global
and local properties of the lens potential.  For example, a
global $m=1$ mode (i.e., non-reflection symmetry) would affect
conclusions about anomalies in direct modeling, but not in our
analysis.  A second reason is that we consider quite general
forms for the lens potential and take care to understand which
generic features affect the cusp relation.  A third point is that
our results are less {\it modeling\/} dependent, less subject to
the intricacies of fitting data and using minimization routines.
A fourth advantage of our analysis is that, rather than simply
showing that standard models fail to fit a lens, it clearly
diagnoses why.  We believe that these benefits go a long way
toward establishing that small-scale structure in lens galaxies
is real and can be understood.

We must address a question that is purely semantic but nevertheless
important:  Where do we draw the line between a normal ``smooth''
lens potential and ``small-scale structure''?  Taking a pragmatic
approach, we consider ``smooth'' to mean any features known to be
common in (early-type) galaxies: certain radial density profiles,
reasonable ellipticities, small octopole modes representing
``disky'' or ``boxy'' isophotes, and reasonable external shears.
We consider ``small-scale structure'' to be anything whose presence
in early-type galaxies would be notable.  This can include stars
--- although stars are obviously abundant in galaxies, detecting
the gravitational effects of individual stars is still interesting
--- and dark matter clumps, which seem to have generated the most
interest.  But it may also include tidal streams, massive or offset
disk components (see Quadri et al.\ 2003; M\"oller et al.\ 2003),
large-amplitude multipole density fluctuations (see Evans \& Witt
2003), etc.  We emphasize that our analysis, or indeed any analysis
that considers only the image positions and broad-band flux ratios
in individual lenses, cannot distinguish between these types of
small-scale structure.  The most general conclusion we can draw
from flux ratio anomalies is that the lens potential contains
structure on scales comparable to or smaller than the separation
between the images.  Further data and analysis is required to
determine the nature of the small-scale structure (e.g., Wo\'zniak
et al.\ 2000; Metcalf 2002; Kochanek \& Dalal 2003; Moustakas \&
Metcalf 2003; Schechter et al.\ 2003; Wisotzki et al.\ 2003).

The layout of the paper is as follows.  We begin in \refsec{configs}
by reviewing quadruple imaging and introducing a way to characterize
4-image configurations quantitatively.  (In this paper we consider
only 4-image lenses.)  In \refsec{ucusps} we discuss cusp image
configurations and present the generic, universal relation that
should be obeyed by the image magnifications for sources near an
ideal cusp.  We then test this ideal relation, first using analytic
results for simple lens potentials (\refsec{simple}), and then
with Monte Carlo simulations of realistic lens populations
(\refsec{monte}).  In \refsec{obs} we apply the cusp relation to
observed lenses, using violations of the relation to identify lenses
that require small-scale structure.  We offer our conclusions in
\refsec{concl}.  Several appendices present supporting technical
material.  In \refapp{analytics} we derive the universal relations
between the image positions and magnifications for sources near an
ideal cusp.  In \refapp{exact} we obtain exact analytic solutions
of the lens equation for two families of lens potentials, which
can be used to obtain exact analytic expressions for the realistic
cusp relation.

\section{Characterizing 4-Image Lenses}
\label{sec:configs}

Nineteen quadruply-imaged lens systems have appeared in the
literature, and they are listed in \reftab{data}.  This count
includes only systems that have exactly four images of a given
source, and where the images appear point-like at some wavelength.
It includes the 10-image system B1933+503, which is complex only
because there are three distinct sources; none of the sources has
an image multiplicity larger than four (Sykes et al.\ 1998).  By
contrast, it excludes PMN~J0134$-$0931 and B1359+154 because they
have multiplicities larger than four due to the presence of multiple
lens galaxies (Rusin et al.\ 2001; Keeton \& Winn 2003; Winn et
al.\ 2003). One other lens, 0047$-$2808, is almost certainly
quadruply-imaged as well (Warren et al.\ 1996, 1999; Koopmans \&
Treu 2003), but its lack of point-like images makes it difficult
to analyze with the usual techniques used for point-like systems.

\begin{figure}[t]
\centerline{\epsfxsize=3.2in \epsfbox{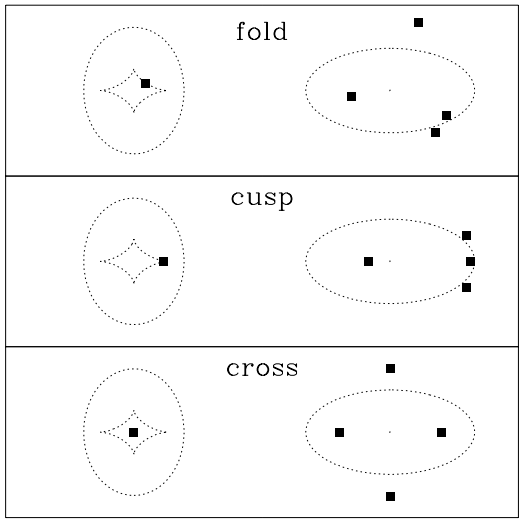}}
\caption{
The three basic configurations of 4-image lenses: fold (top), cusp
(middle), and cross (bottom). In each panel, the figure on the left
shows the caustics and source position in the light source plane,
while the figure on the right show the critical curves and image
positions in the image plane.
}\label{fig:quadtypes}
\end{figure}

Mathematically, quadruple imaging can be described in terms of the
critical curves and caustics of the lens potential.  (See the
monographs by Schneider et al.\ 1992 and Petters et al.\ 2001 for
thorough reviews of lens theory.)  Critical curves are curves in the
image plane where the lensing magnification is formally infinite,
and caustics are the corresponding curves in the light source plane.
The properties of these curves can be studied with catastrophe
theory; for our purpose the important result is that the
astroid-shaped caustic that is associated with quadruple imaging
has a generic shape that leads to three generic configurations of
4-image lenses (see \reffig{quadtypes}).  Sources near a cusp in the
caustic produce ``cusp'' configurations with three of the images
lying close together on one side of the lens galaxy.  Source near
the caustic but not near a cusp produce ``fold'' configurations with
two of the images lying close together.  Sources not close to the
caustic produce relatively symmetric ``cross'' configurations.

\begin{figure}[t]
\centerline{\epsfxsize=2.1in \epsfbox{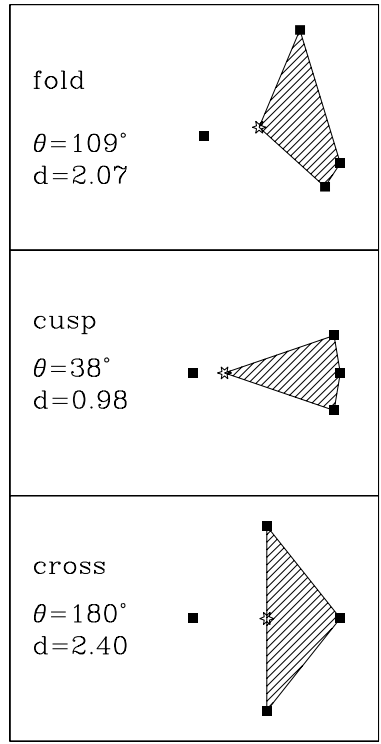}}
\caption{
Sample image triplets for the image configurations from
\reffig{quadtypes}, together with the values of the opening angle
$\theta$ and the separation $d$ (in units of the Einstein radius of
the lens).  The star shows the position of the lens galaxy in each
system.
}\label{fig:samptrip}
\end{figure}

Although it may seem easy to label an observed lens as a fold, cusp,
or cross, the categories actually blend together so it is important
to develop a more quantitative way to characterize image configurations.
To quantify a triplet of images (as in a cusp configuration), let $d$
be the maximum separation between the three images, and let $\theta$
be the opening angle of the polygon spanned by the three images,
measured from the position of the lens galaxy.  Each 4-image lens has
four distinct triplets and hence four values of $\theta$ and $d$.  We
can identify image triplets associated with cusps as those where
$\theta$ and/or $d$ is small (see \reffig{samptrip}).  Even though
there is no rigorous definition of when $\theta$ and $d$ are ``small''
enough to indicate a cusp configuration, we shall see below that
these are useful quantities for characterizing the range of image
configurations.

\section{Universal Magnification Relation for Cusps}
\label{sec:ucusps}

In this section we briefly review the lensing of a source close to
and inside an ideal cusp and present the magnification relation
used in our analysis. \refapp{analytics} discusses lensing near a
cusp in considerably more detail, and presents additional position
and magnification relations for cusp images.  This analysis applies
to ordinary cusps; it may not be valid for so-called ramphoid cusps
(or cusps of the second kind), but such cusps have not been observed
and are expected to be rare in lensing situations of astrophysical
interest (see Petters \& Wicklin 1995; Oguri et al.\ 2003).

In the vicinity of a cusp, the lens equation relating the source
position $\bu$ to the image position $\bt$ can be written to third
order in $\bt$ as a polynomial mapping,
\begin{equation} \label{eq:le-cusp0}
  u_1 = c \ \theta_1 \  + \  \frac{b}{2}\  \theta^2_2 \, , \qquad
  u_2 = b \  \theta_1 \  \theta_2 + a \  \theta_2^3 \, .
\end{equation}
The coordinates $\bu$ and $\bt$ are local orthogonal coordinates
that are related to the global coordinates $\bby$ and $\bx$ of the
lens system by $\bu \equiv {\rm M} \bby$ and
$\bt \equiv {\rm M}\bx$, where the transformation matrix ${\rm M}$
depends on the lens potential.  For the simple cases that we study
in \refsec{simple} and \refapp{exact}, ${\rm M}$ is the identity
matrix and the $\bt$ and $\bu$ coordinate systems are simply the
$\bx$ and $\bby$ coordinate systems translated so the cusp point
is at the origin.  The constant coefficients $a$, $b$, and $c$ are
given by derivatives of the potential at the critical point (see
eq.~\ref{eqn:coeff} in \refapp{analytics}).

Solving for $\theta_1$ in the left-hand side of \eq{le-cusp0} and
substituting into the right-hand side, one obtains a cubic equation
for $\theta_2$ that depends on $a$, $b$, $c$, and the source position
$\bu$.  Inside the caustic, there are three real solutions to this
cubic equation, and thus three images of the source.  It is possible
to derive six independent relations between the positions and
magnifications of these images.  Unfortunately, only one of these
relations can be recast to depend only on directly observable
properties: the well-known magnification sum rule (Schneider \&
Weiss 1992; Zakharov 1995; Petters et al.\ 2001, p.~339),
\begin{equation}
  \mu_1 + \mu_2 +  \mu_3 = 0 \, ,
\label{eq:univ-mag-rule-main}
\end{equation}
where the $\mu_i$ are the signed magnifications of the three images.
The other relations depend on properties that are not directly
observable, such as the position of the source or the mapping
coefficients $a$, $b$, and $c$.

\section{The Cusp Relation in Simple Lens Potentials}
\label{sec:simple}

The derivation of the ideal cusp relation \eq{univ-mag-rule-main}
relies on the assumption that the lensing map has the polynomial
form of \eq{le-cusp0}.  Since this form is a truncated Taylor
series expansion near the cusp point, we should expect the cusp
relation to be exact only for sources asymptotically close to the
cusp.  In this section we begin to quantify the deviation from the
ideal cusp relation that arise from the higher order terms in the
lensing map, using simple examples to illustrate the effects of
the radial profile, ellipticity, shear, and multipole perturbations
of the lens potential.

The magnifications appearing in the cusp relation are not directly
observable, but we can follow Mao \& Schneider (1998) and divide
out the unknown source flux by defining the dimensionless quantity
\begin{equation}
  \Rcusp \equiv \frac{|\mu_1+\mu_2+\mu_3|}{|\mu_1|+|\mu_2|+|\mu_3|}
  = \frac{|F_1+F_2+F_3|}{|F_1|+|F_2|+|F_3|}\ ,
\end{equation}
where the $\mu_i$ are the magnifications and the $F_i$ the observed
fluxes, both with signs indicating the image parities. The parities
can be determined unambiguously because in any triplet of adjacent
images, the two outer images have the same parity while the middle
image has the opposite parity (see Schneider et al.\ 1992; Petters
et al.\ 2001).  The ideal cusp relation has the form $\Rcusp=0$. 

Note that we have defined $\Rcusp$ to be non-negative.  Several
recent studies (Schechter \& Wambsganss 2002; Keeton 2003; Kochanek
\& Dalal 2003) have pointed out that small-scale structure tends to
suppress negative-parity images more often than it amplifies
positive-parity images, while global perturbations generally do not
distinguish between images with different parities.  In an ensemble
of lenses with flux ratio anomalies, skewness in the {\it signed\/}
$\Rcusp$ distribution may therefore distinguish local from global
perturbations.  However, the statistical nature of this argument
precludes its use in individual lenses.  Since we seek a method of
identifying anomalies in individual lenses, we consider only the
unsigned quantity.

We first study the cusp relation analytically using two families
of lens potentials where it is possible to obtain exact solutions
of the lens equation.  In one family, the galaxy is assumed to be
spherical but is allowed to have a general power law surface density
profile $\Sigma \propto r^{\alpha-2}$ and to have an external shear
$\gamma$.  In the other family, the galaxy is assumed to have an
``isothermal'' profile $\Sigma \propto r^{-1}$ but is allowed to
have a complex angular structure, including shear; we specifically
consider an ellipsoidal galaxy perturbed by multipole density
fluctuations.  \refapp{exact} describes the two families of lens
potentials in detail and gives solutions for the positions and
magnifications of images corresponding to sources on a symmetry axis
of the lens potential.

\begin{figure}[t]
\centerline{\epsfxsize=3.4in \epsfbox{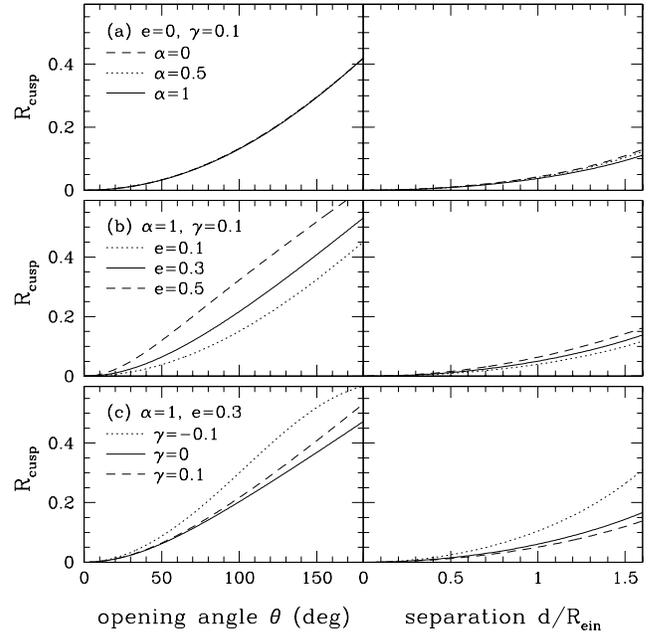}}
\caption{
The cusp relation residual $\Rcusp$ as a function of the opening
angle $\theta$ and the separation $d$ of an image triplet, plotted
for various lens potentials using the analytic solutions to the lens
equation for sources on the major axis of the potential.  In panel
(c), $\gamma>0$ ($\gamma<0$) represents a shear aligned with
(orthogonal to) the major axis of the galaxy.
}\label{fig:anal1}
\end{figure}

\reffig{anal1} shows $\Rcusp$ versus the opening angle $\theta$ and
separation $d$ of an image triplet, for various potentials with
different radial profiles, ellipticities, and shears.  In general,
$\Rcusp$ is small when $\theta$ and $d$ are small (indicating that
the source is very near a cusp), and grows as $\theta$ and $d$ grow
(indicating that the source is moving farther from the cusp).  The
analytic results allow us to understand how departures from the ideal
cusp relation depend on properties of the lens potential.  We see that
radical changes in the radial profile of the lens potential --- from
$\alpha=1$ (isothermal) to $\alpha=0$ (point mass) --- have a negligible
effect on the cusp relation.  By contrast, moderate changes in the
ellipticity and shear can affect the cusp relation by tens of percent.
The fact that the cusp relation is quite sensitive to ellipticity,
moderately sensitive to shear, and not very sensitive to the radial
profile makes sense: reasonable changes in the angular structure of
the potential ($e$ and $\gamma$) can affect nearby images quite
differently, while reasonable changes in the radial profile cannot.
Incidentally, we note that when considering fixed ellipticity and
shear amplitudes, $\Rcusp$ can be larger when the two are orthogonal
than when they are aligned.

\begin{figure}[t]
\centerline{\epsfxsize=2.2in \epsfbox{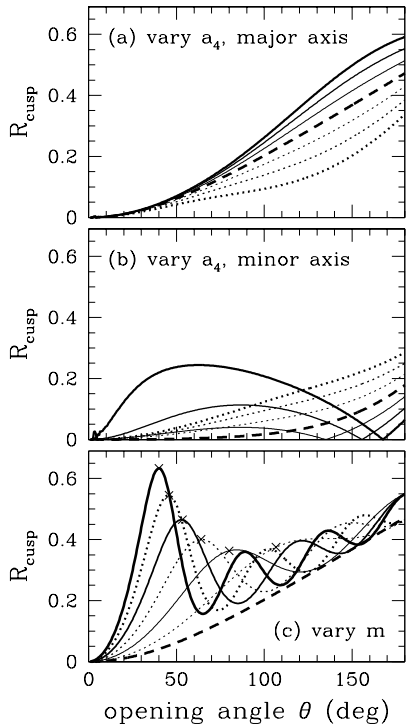}}
\caption{
Effects of multipole perturbations on the cusp relation, for
isothermal ellipsoid lens potentials with $e=0.3$ and $\gamma=0$.
The heavy dashed curves all show a reference case with no multipole
modes.
(a) The effects of $m=4$ density fluctuations with amplitude
$a_4 = (\pm0.01,\pm0.02,\pm0.03)$ indicated by increasing line
thickness.  Solid (dotted) curves correspond to $a_4>0$ ($a_4<0$).
The sources lie on the major axis of the lens potential.
(b) Similar to (a), but for sources on the minor axis of the
potential
(c) The effects of multipole perturbations of different orders,
all with amplitude $a_m=0.02$.  Lines of increasing thickness and
alternating type indicate $m=(6,8,10,12,14,16)$.  The cross on each
curve marks the point with $\theta=640\arcdeg/m$.  The sources lie
on the major axis of the lens potential.
}\label{fig:anal2}
\end{figure}

The effects of multipole density perturbations are shown in
\reffig{anal2}, for lens potentials with an ``isothermal''
($\alpha=1$) radial profile.  Multipole modes with $m=3$ or 4 and
amplitudes of a few percent are common in the isophotes of observed
early-type galaxies (Bender et al.\ 1989; Saglia et al.\ 1993; Rest
et al.\ 2001) and in the isodensity contours of simulated galaxy
merger remnants (Heyl, Hernquist, \& Spergel 1994; Naab \& Burkert
2003; Burkert \& Naab 2003); in particular, $m=4$ modes with
amplitudes $a_4>0$ can represent small disk-like components in
bulge-dominated galaxies, which are not unusual (Kelson et al.\
2000; Tran et al.\ 2003).  Such modes might have a significant
effect on the magnifications of lensed images (Evans \& Witt
2002; M\"oller et al.\ 2003).  We find that $m=4$ modes do not
significantly increase $\Rcusp$ for cusp triplets with
$\theta \lesssim 90\arcdeg$ when the source is on the major axis
of the lens potential (\reffig{anal2}a).  However, they can
create remarkably large values of $\Rcusp$ even for small
$\theta$ when the source is on the minor axis (\reffig{anal2}b).
At fixed amplitude, higher order modes produce progressively larger
values of $\Rcusp$ at smaller angles (\reffig{anal2}c).\footnote{For
the high-order multipole modes we do not show sources on the minor
axis of the potential, because on the minor axis the caustics often
have complicated butterfly catastrophes that need not satisfy the
cusp relation (see \refapp{GISO}).}  The position of the peak in
the $\Rcusp$ curve for different values of $m$ can be approximated
as $\theta_{\rm peak} \sim 640\arcdeg/m$.  This result describes
our fiducial case with $e=0.3$ and $a_m=0.02$; varying $e$ and
$a_4$ has a small ($\lesssim$5\%) effect on the position of the
peak, but a large effect on the amplitude of the peak.  Thus, we
can say as a rule of thumb that image triplets with angle $\theta$
are significantly affected only by modes with
$m \gtrsim 640\arcdeg/\theta$.  We conclude that it is important to
consider multipole effects in the cusp relation analysis.  But as it
is not clear that real galaxies have percent-level perturbations in
modes beyond $m \approx 4$, it is equally important to hold the
perturbations to reasonable levels.

\begin{figure}[t]
\centerline{\epsfxsize=3.4in \epsfbox{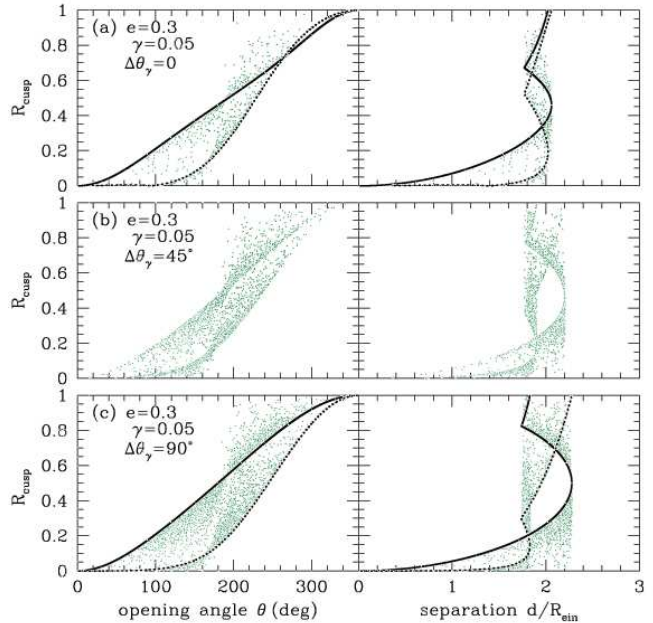}}
\caption{
$\Rcusp$ versus $\theta$ and $d$, for various values of the
ellipticity and shear and the angle $\Delta\theta_\gamma$ between
them.  The points show results for Monte Carlo simulations of random
source positions.  The heavy curves show the analytic results for
on-axis sources, where solid (dotted) curves indicate the major
(minor) axis of the lens potential.  Analytic results are available
only for $\Delta\theta_\gamma=0$ and $90\arcdeg$.
}\label{fig:scatt}
\end{figure}

So far we have studied only sources lying on a symmetry axis of the
lens potential.  For the more general case we turn to Monte Carlo
simulations.  We pick random source positions and solve the lens
equation (using the algorithm and software by Keeton 2001b) to
generate a catalog of mock lenses.  We compute $\theta$, $d$, and
$\Rcusp$ for each triplet in each 4-image lens, and then plot
$\Rcusp$ versus $\theta$ or $d$ for all triplets.   \reffig{scatt}
shows sample results for isothermal ellipsoid galaxies with shear.
The most important result is that over the region of interest for
the cusp relation ($\theta \lesssim 180\arcdeg$ and
$d/\Rein \lesssim 1.7$) there is a firm upper envelope on the values
of $\Rcusp$.\footnote{The break at $d/\Rein \sim 1.7$ is simple to
understand.  This separation corresponds to an image triplet
comprising an equilateral triangle inscribed within the Einstein
ring.  When the separation reaches this value the images are so
spread out that they can no longer be associated with a cusp.}  In
fact there are two envelopes: one each for major and minor axis
cusps.  Moreover, in lenses with reflection symmetry the envelope
corresponds to sources on the symmetry axis.  To understand this
result, in \refapp{PL} we prove that $\Rcusp$ is a local maximum
on the symmetry axis of an isothermal sphere plus shear.  Messy
algebra hinders a rigorous analysis of other potentials, but
intuition and the Monte Carlo simulations suggest that the result
is generally true.  In other words, the analytic results for
on-axis sources provide a simple and important upper bound on
$\Rcusp$.

To summarize, the ideal cusp relation breaks down for sources a
small but finite distance from the cusp, but in a way that can be
understood and quantified.  The realistic cusp relation is mainly
sensitive to the angular structure of the lens potential, not the
radial profile.  The important quantities are the ellipticity, shear,
and strength of multipole density fluctuations.  For the subset of
cusps that possess a symmetry axis, sources on that axis provide a
strict upper bound on $\Rcusp$ over the interesting range of $\theta$
and $d$ that can often be derived analytically.

\section{The Cusp Relation in Realistic Lens Populations}
\label{sec:monte}

If we knew the ellipticity, shear, and multipole perturbations for
individual observed lenses, we could use the previous analysis to
compute how much $\Rcusp$ can deviate from zero for smooth potentials
and then conclude that larger values represent flux ratio anomalies.
Unfortunately, the three key quantities are not directly observable.
The ellipticity and multipole perturbations of the mass need not be
the same as those of the light (e.g., Keeton, Kochanek \& Falco
1998), and in any case the shear cannot be directly observed.  The
three quantities could be constrained with lens models, but we seek
to avoid explicit modeling to the extent possible.  Instead, our
approach is to adopt observationally-motivated priors on the
ellipticity, multipole perturbations, and shear, and use Monte
Carlo simulations to obtain a sample of realistic lens potentials
and derive probability distributions for $\Rcusp$.  In this section
we describe the priors (\refsec{monte-input}) and methods
(\refsec{monte-method}) for the simulations.

\subsection{Input distributions}
\label{sec:monte-input}

We consider only early-type galaxies, because they are expected
to dominate the lensing optical depth due to their large average
mass (e.g., Turner, Ostriker, \& Gott 1984; Fukugita \& Turner
1991).  Indeed, $\sim$80--90\% of observed lens galaxies have
properties consistent with being massive ellipticals (Keeton et
al.\ 1998; Kochanek et al.\ 2000; Rusin et al.\ 2003a).  The
distinction between ellipticals and spirals is important, because
disk-dominated galaxies that are viewed close to edge-on can
produce cusp configurations that deviate significantly from the
cusp relation (Keeton \& Kochanek 1998).  Several of the lenses
for which we identify flux ratio anomalies are confirmed ellipticals,
and none of them have properties suggesting that they are spirals
(Impey et al.\ 1996; Burud et al.\ 1998; Fassnacht et al.\ 1999;
Jackson et al.\ 2000; Inada et al.\ 2003; Sluse et al.\ 2003;
Rusin et al.\ 2003a).\footnote{Despite a suggestion by M\"oller
et al.\ (2003) that the lens galaxy in B2045+265 might be a 
spiral, its structural and dynamical properties are fully consistent
with being an elliptical (Rusin et al.\ 2003a), and no disk-like
structure is evident in Hubble Space Telescope images (C.~Kochanek,
private communication).}  

We allow the simulated galaxies to have ellipticity and also octopole
($m=4$) perturbations, with distributions drawn from observations of
isophote shapes in early-type galaxies.  Even if the shapes of the
mass and light distributions are not correlated on a case-by-case
basis, it seems likely that their distributions are similar (see
Rusin \& Tegmark 2001 for a discussion).  Indeed, the distribution of
isodensity contour shapes in simulated merger remnants is very similar
to the observed distribution of isophote shapes (Heyl et al.\ 1994;
Naab \& Burkert 2003; Burkert \& Naab 2003).  Multipole perturbations
beyond $m \ge 5$ have generally not been reported, but it is likely
that they must have relatively low amplitudes to be compatible with
observations.  Lower-order $m=3$ modes have been reported with
amplitudes comparable to $m=4$ modes (e.g., Rest et al.\ 2001), but
we do not consider them here because they are not reported in the
samples we use, and because at fixed amplitude higher-order modes
produce larger deviations in the cusp relation (see \reffig{anal2}).
Our approach is formally equivalent to studies that explicitly
include a disk-like mass component (e.g., M\"oller et al.\ 2003),
since small disks can be treated as $m=4$ multipole perturbations.
We use only isothermal galaxies, because the radial profile of the
lens galaxy does not significantly affect the cusp relation.  The
galaxy mass is unimportant, because it simply sets a length scale
(the Einstein radius) that can be scaled out by using the dimensionless
separation $d/\Rein$.

\begin{figure}[t]
\centerline{\epsfxsize=3.4in \epsfbox{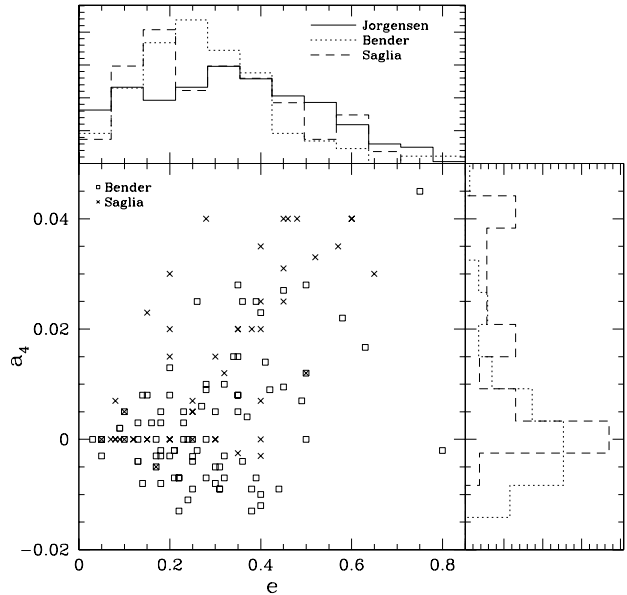}}
\caption{
The main panel shows the $e$ and $a_4$ values for the galaxies in the
Bender et al.\ (1989) and Saglia et al.\ (1993) samples.  Galaxies
with $a_4>0$ ($a_4<0$) have disky (boxy) isophotes.  The right panel
shows the histograms of $a_4$.  The top panel shows histograms of
$e$ for these samples and also the J{\o}rgensen et al.\ (1995) sample. 
}\label{fig:compare}
\end{figure}

We use ellipticity and octopole distributions from three different
samples.  J{\o}rgensen et al.\ (1995) report ellipticities for 379
E and S0 galaxies in 11 clusters, including Coma.  Their ellipticity
distribution has mean ${\bar e}=0.31$ and dispersion $\sigma_e=0.18$.
Since the J{\o}rgensen et al.\ sample does not include octopole
amplitudes, we consider two smaller samples that do.  Bender et
al.\ (1989) report ellipticities and octopoles for 87 nearby,
bright elliptical galaxies.  Their ellipticity distribution has
${\bar e}=0.28$ and $\sigma_e=0.15$, while their octopole distribution
has mean ${\bar a}_4=0.003$ and dispersion $\sigma_{a_4}=0.011$.
Finally, Saglia et al.\ (1993) report ellipticities and octopole
amplitudes for 54 ellipticals in Coma, with ${\bar e}=0.30$ and
$\sigma_e=0.16$, and ${\bar a}_4=0.014$ and $\sigma_{a_4}=0.015$.
Compared to the Bender et al.\ sample, the Saglia et al.\ sample
has a higher incidence of galaxies with strong disky perturbations
($a_4>0$).  Considering all three samples allows us to examine
whether our conclusions depend systematically on the input $(e,a_4)$
distributions (although we note that since the J{\o}rgensen et al.\
and Saglia et al.\ samples both include galaxies in Coma they are
not fully independent).  \reffig{compare} shows the different samples
and suggests that $e$ and $a_4$ are correlated such that highly
elliptical galaxies tend to have significant disky perturbations.
Our analysis includes this correlation explicitly by using the
observed joint distribution of $e$ and $a_4$.

For the shear amplitude we adopt a lognormal distribution with
median $\gamma=0.05$ and dispersion $\sigma_\gamma=0.2$ dex.  This
is consistent with the distribution of shears expected from the
environments of early-type galaxies, as estimated from $N$-body
and semi-analytic simulations of galaxy formation (Holder \&
Schechter 2003). It is broadly consistent with the empirical
distribution of shears required to fit observed lenses, when
selection biases related to the lensing cross section and
magnification bias are taken into account (see Holder \& Schechter
2003).  The mean shear is also consistent with the typical value
needed to explain misalignments between the light and mass in
observed lenses (Kochanek 2002).  We assume random shear
orientations.

\subsection{Simulation methods}
\label{sec:monte-method}

\begin{figure*}[t]
\centerline{
  \epsfxsize=3.3in \epsfbox{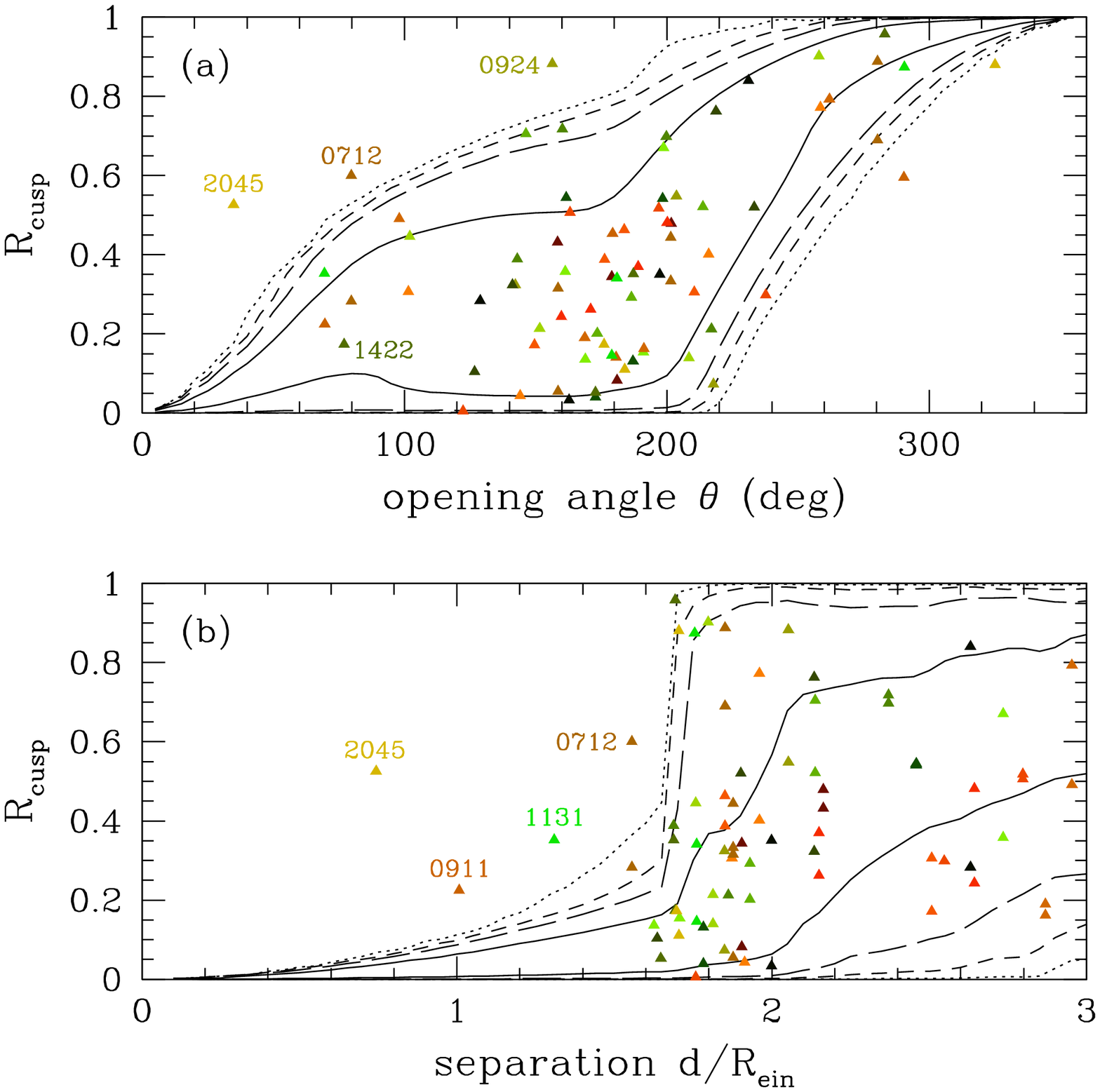}
  \epsfxsize=3.3in \epsfbox{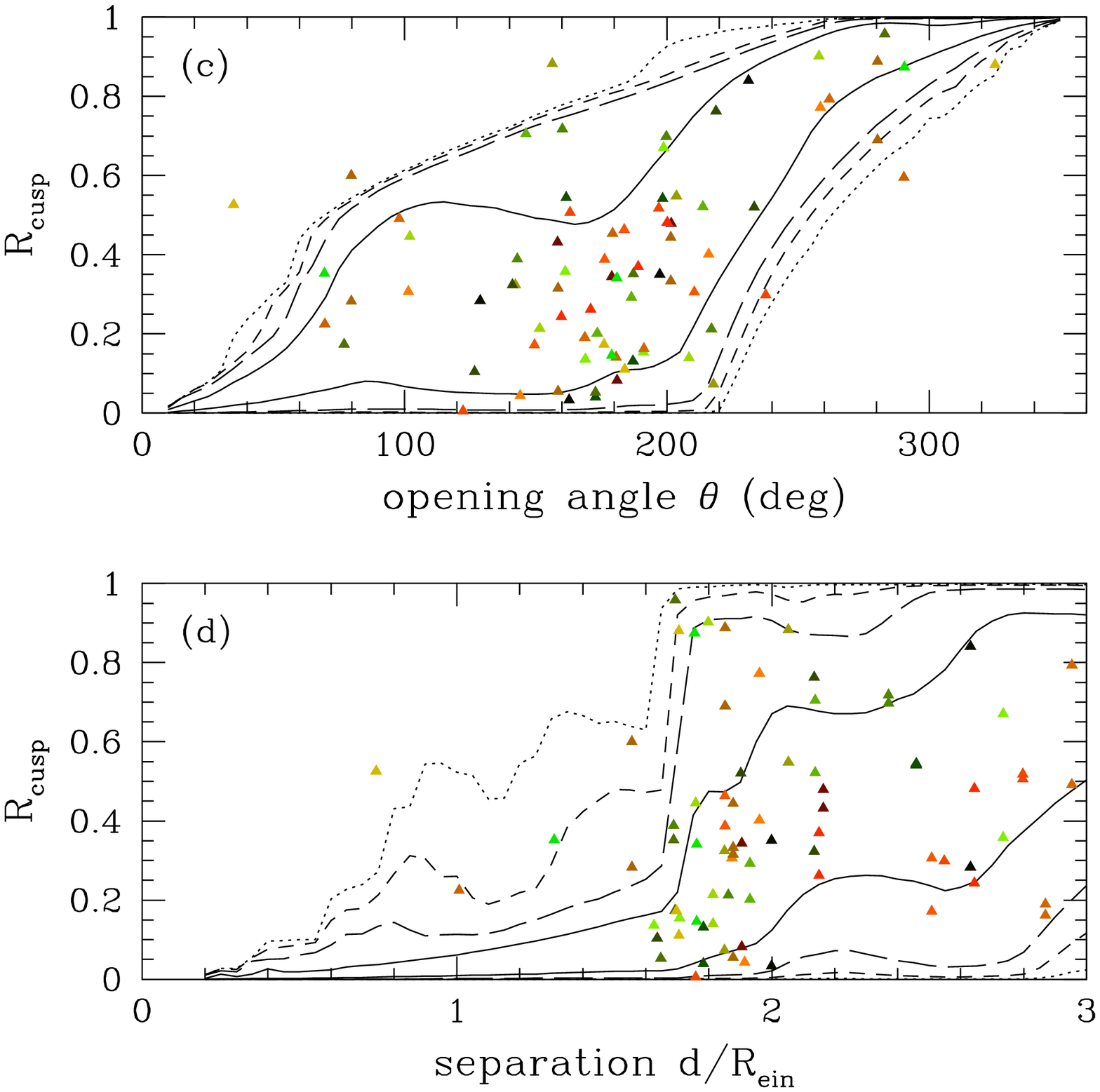}
}
\caption{
The curves show contours of constant conditional probability
$p(\Rcusp|\theta)$ (top) and $p(\Rcusp|d)$ (bottom), from Monte Carlo
simulations using the J{\o}rgensen et al.\ input data (left) or the
Bender et al.\ input data (right).  The contours are drawn at the
68\% (solid), 95\% (long-dashed), 99\% (dashed), and 99.9\% (dotted)
confidence levels.  The points show the observed values of $\Rcusp$
for the known 4-image lenses.  B0712+472 appears twice; the lower
value of $\Rcusp$ corresponds to the radio data, while the higher
value (labeled) corresponds to the optical/near-IR data (see text).
}\label{fig:Rcusp-obs}
\end{figure*}

We use each input distribution to run a large Monte Carlo simulation
containing $\sim\!10^{6}$ 4-image lenses.  With the J{\o}rgensen et
al.\ sample we draw 2000 ellipticities from the observed ellipticity
distribution and give each a random shear.  With the Bender et al.\
and Saglia et al.\ samples we use only the observed $(e,a_4)$ pairs
to make sure we include the apparent correlation between the two
quantities, but we use each pair with 100 different random shears;
thus we consider 8700 and 5400 lens potentials for the Bender et al.\
and Saglia et al.\ samples, respectively, but we need to remember
that these represent only 87 and 54 different ellipticity and
octopole measurements.  For each potential, we pick random sources
with a uniform density of $\sim\!10^{3}\,\Rein^{-2}$ in the source
plane and solve the lens equation using the algorithm and software
by Keeton (2001b).  We have verified that our results are not
sensitive to the number of shears and density of sources used.

To understand how our mock lenses compare to observed samples it
is important to consider two selection effects.  First, the cross
section for 4-image lenses is very sensitive to ellipticity and shear,
but our uniform sampling of the source plane ensures that each lens
potential is {\it automatically\/} weighted by the correct cross
section.  Second, magnification bias can favor lenses with higher
amplifications.  While this effect is important when comparing 4-image
lenses to 2-image lenses (e.g., Keeton et al.\ 1997; Rusin \& Tegmark
2001), it is much less important when comparing different 4-image
lenses against each other.  If anything, it would favor the sources
very near a cusp or fold that yield extremely magnified lenses that
best satisfy the cusp/fold relations, giving more weight to lenses
with {\it smaller\/} deviations from the ideal relations.  We
therefore neglect magnification bias and believe that this is a
conservative approach.

We compute $\theta$, $d$, and $\Rcusp$ for each image triplet in
each mock 4-image lens, and use this ensemble to determine the
conditional probability distributions $p(\Rcusp|\theta)$,
$p(\Rcusp|d)$, and $p(\Rcusp|d,\theta)$ --- the probability of having
a particular value of $\Rcusp$ given $\theta$, $d$, or both.  For
example, \reffig{Rcusp-obs} shows curves of constant conditional
probability $p(\Rcusp|\theta)$ and $p(\Rcusp|d)$ versus $\theta$ and
$d$.  This figure is basically a modified version of \reffig{scatt}
where we have averaged over appropriate ellipticity, octopole, and
shear distributions.  It is interpreted as saying that 68\% of
triplets in our sample of mock lenses lie in the region between
the solid curves, 99\% lie between the dashed curves, and so forth.
To the extent that our simulations encompass the range of
ellipticities, octopoles, and shears in real populations of
early-type galaxies, we can conclude that any points lying outside
the contours represent flux ratios that are inconsistent with smooth
lens potentials.

\section{Application to Observed Lenses}
\label{sec:obs}

We can now use our theoretical analysis to evaluate observed lenses,
seeking to identify systems that violate the cusp relation and
therefore have anomalous flux ratios.  We first summarize the data
(\refsec{obs-data}) and then present our results (\refsec{obs-results}).

\subsection{Data}
\label{sec:obs-data}

The data for the nineteen published 4-image lenses are given in
\reftab{data}.  Only five of the lenses are thought to have cusp
configurations, but we can still apply the cusp relation analysis
to all of them to see what we learn.  In the table we list the four
different image triplets for each lens, with the opening angle
$\theta$, the image separation $d$, and the observed value of
$\Rcusp$ for each.  If the lens galaxy position is known, the angle
$\theta$ is fully determined by the data; if not, we estimate
$\theta$ using the galaxy position estimated from lens models,
which is a fairly model-independent prediction.  The separation $d$
is determined directly from the data.  To normalize it we need the
Einstein radius $\Rein$, which must be derived from a lens model
but is insensitive to the assumed model; different models generally
yield the same Einstein radius with systematic uncertainties of
just a few percent (Cohn et al.\ 2001; Rusin et al.\ 2003b).

When we measure $\Rcusp$ we need to consider systematic uncertainties
due to effects like source variability and the lens time delay, scatter
broadening (at radio wavelengths), and differential extinction by
patchy dust in the lens galaxy (at optical wavelengths).  (Extinction
by dust in our own Galaxy does not affect the flux ratios, because
it affects all images equally.)  Dalal \& Kochanek (2002) advocate
adopting a fiducial estimate of 10\% uncertainties in the flux ratios
to account for these effects, but this is likely to be quite
conservative.  For most lenses the uncertainties are irrelevant
because the measured values of $\Rcusp$ lie well within the expected
distribution, so for \reftab{data} we use 10\% flux uncertainties for
simplicity.  We want to be more careful about the error budgets for
lenses suspected of having flux ratio anomalies, so we discuss them
individually in the next section.

For comparison, \reftab{data} also gives values for $\Rcusp$
predicted by standard lens models consisting of an isothermal
ellipsoid with an external shear.\footnote{The only exception is
B1555+375, where the ellipsoid plus shear model is somewhat ambiguous
and we use models with a slightly different parameterization of
the quadrupole moment of the lens potential; see Turner et al.\
(C.~Turner, C.~R.~Keeton, \& C.~S.~Kochanek, in prep.) for technical
details.}  Only the image positions (not the flux ratios) were used
as constraints.  In MG~0414+0534, RX~J0911+0551, and B1608+656 the
lens models also include the perturbative effects of an observed
satellite galaxy near the main lens galaxy.  We stress that the
lens models are not actually used in seeking flux ratio anomalies
(other than for estimating $\Rein$, as discussed above).  They are
included only as a general indication of what to expect for $\Rcusp$
from smooth lens models for these systems.

\subsection{Results}
\label{sec:obs-results}

\reffig{Rcusp-obs} shows the observed values of $\Rcusp$ superposed
on the predicted confidence contours.  The five cusp lenses can be
identified as the points with $\theta \lesssim 80\arcdeg$ (B0712+472
appears twice: once for radio data, and once for optical/near-IR
data).  Most of the observed lenses lie within the predicted confidence
region, so according to this analysis they are not obviously inconsistent
with smooth lens potentials.  We note that the predicted confidence
contours are very similar for the J{\o}rgensen et al.\ and Bender et
al.\ input data (and also for the Saglia et al.\ input data, not shown).
The main difference is that the presence of octopole perturbations in
the Bender et al.\ input data causes the confidence contours to
stretch to higher values of $\Rcusp$ at $d/\Rein \lesssim 1.7$, which
is what we expect from the theoretical analysis in \refsec{simple}.
(The contours for the Bender et al.\ input data are somewhat noisy
due to the relatively small number of ellipticity and octopole
measurements.)  Thus, contrary to the claim by M\"oller et al.\
(2003), we find that adding (properly-normalized) disky components
to elliptical galaxies does not have an enormous effect on the cusp
relation.  We shall explain below which of our conclusions are or
are not affected by the presence of octopole terms, or more generally
by changes in the input data.

Several of the lenses are obvious outliers.  The cusp lens B2045+265
lies outside all contours.  The cusp lens B0712+472 lies outside all
$p(\Rcusp|\theta)$ contours and either outside or just inside the
99.9\% confidence contour for $p(\Rcusp|d)$, depending on the input
data.  The cusp lenses 1RXS~J1131$-$1231 and RX~J0911+0551 stand out
relative to $p(\Rcusp|d)$ for the J{\o}rgensen et al.\ input data
but not for the Bender et al.\ (or Saglia et al.) input data.
(Incidentally, RX~J0911+0551 and B2045+265 are also responsible for
the $p(\Rcusp|\theta)$ outliers at $\theta=290\arcdeg$ and
$\theta=325\arcdeg$, respectively.)  The fifth cusp lens B1422+231
does not stand out in this analysis.  Finally, the lens
SDSS~J0924+0219 is an outlier with respect to $p(\Rcusp|\theta)$
even though it is not a cusp configuration.

\begin{figure}[t]
\centerline{\epsfxsize=3.3in \epsfbox{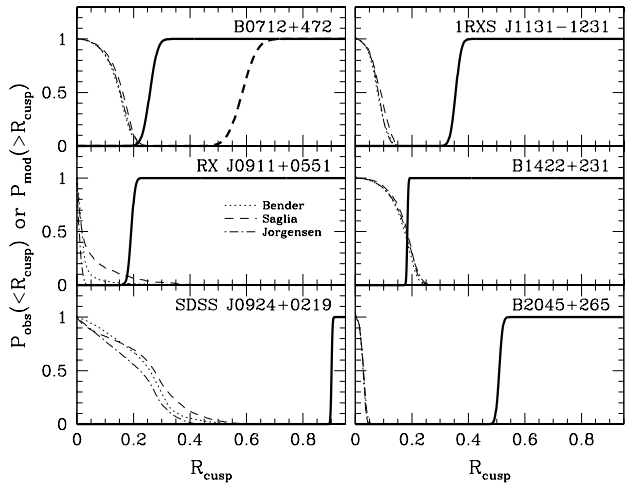}}
\caption{
Cumulative distributions for $\Rcusp$ for the six lenses discussed
individually.  The rising curves show the probability
$P_{\rm obs}(<\!\Rcusp)$ that the measurement of $\Rcusp$ is smaller
than some value; for B0712+472 the solid (dashed) curves denote the
radio (optical/near-IR) measurements.  The falling curves show the
probability $P_{\rm mod}(>\!\Rcusp|d,\theta)$ that smooth lens
potentials produce $\Rcusp$ larger than some value, with different
line types denoting different input data for the Monte Carlo
simulations.
}\label{fig:joint}
\end{figure}

The joint conditional probability distribution $p(\Rcusp|d,\theta)$
provides an even more powerful way to identify outliers.
\reffig{joint} compares the cumulative probability
$P_{\rm mod}(>\!\Rcusp|d,\theta)$ that smooth potentials produce
$\Rcusp$ larger than some value versus the cumulative probability
$P_{\rm obs}(<\!\Rcusp)$ that the measurement of $\Rcusp$ is smaller
than some value, for the six lenses just mentioned.  The measured
value of $\Rcusp$ is compatible with smooth potentials only if the
curves have a significant overlap.  B2045+265, 1RXS~J1131$-$1231,
and SDSS~J0924+0219, and B0712+472(optical) are clear outliers;
B0712+472(radio) and RX~J0911+0551 are marginal outliers; and
B1422+231 is the only case where the observed and predicted
distributions are clearly compatible.  We now discuss each of these
lenses individually.

\subsubsection{B2045+265}

Fassnacht et al.\ (1999) give eight measurements of the radio
fluxes for B2045+265 at 1.4, 5, 8.5, and 15~GHz, from different
radio arrays with different resolutions. The mean value and scatter
in $\Rcusp$ is $0.516\pm0.018$; the scatter is only slightly larger
than the uncertainty that would be inferred from the quoted flux
errors.  The fact that the $\Rcusp$ values from diverse radio
datasets are consistent within the errors argues against any
significant non-gravitational effects (e.g., scattering).  Also, for
a cusp triplet the time delays are expected to be very short ---
predicted to be $\lesssim$6 hours for B2045+265, and similarly short
for the other cusp lenses --- so they should have no effect on the
measured value of $\Rcusp$.  We therefore believe that $\pm0.018$
represents a reasonable estimate of the uncertainty.  Koopmans et
al.\ (2003a) present 41 measurements of B2045+265 at 5~GHz.  Although
they observe variability that they attribute to scintillation, they
find a mean and scatter in $\Rcusp$ of $0.501\pm0.035$ in excellent
agreement with the value from the Fassnacht et al.\ (1999) data. 

The CfA/Arizona Space Telescope Lens Survey (CASTLES; C.~Kochanek
et al., private
communication)\footnote{See {\tt http://cfa-www.harvard.edu/castles}.}
provides data at optical and near-IR wavelengths from Hubble Space
Telescope imaging.  Their data for B2045+265 yield $\Rcusp=0.501\pm0.037$
in V-band, $0.531\pm0.035$ in I-band, and $0.502\pm0.015$ in H-band.
The colors of the images, and the fact that $\Rcusp$ remains constant
over a factor of three in optical/near-IR wavelength, indicate that
there is little or no differential extinction between the images.
The weighted average of the optical/near-IR data yields
$\Rcusp=0.506\pm0.013$; the excellent agreement with the radio data
suggests that the measured value of $\Rcusp$ is robust and independent
of wavelength, and that the small inferred uncertainties on $\Rcusp$
are realistic.  The weighted average of all measurements is
$\Rcusp=0.509\pm0.010$.

\reffigs{Rcusp-obs}{joint} show that the existence of a flux ratio
anomaly in B2045+265 is beyond doubt.  Image B is simply much too
faint to be consistent with smooth lens potentials, no matter which
input ellipticity and octopole distributions are used.  Attempting
to explain the value of $\Rcusp$ with multipole perturbations would
require a significant amplitude in a mode with $m \approx 16$ (see
\reffig{anal2}).

\subsubsection{B0712+472}

Jackson et al.\ (1998) give three different measurements of the
radio fluxes for B0712+472 at 5~GHz and one measurement at 15~GHz.
The mean and scatter in the value of $\Rcusp$ from the four datasets
is $0.261\pm0.031$.  The data from 41 measurements at 5~GHz by
Koopmans et al.\ (2003a) yield $\Rcusp=0.255\pm0.030$, in excellent
agreement with the Jackson et al.\ (1998) value.  The weighted
average of these measurements is $0.258\pm0.022$.

The optical and near-IR data from CASTLES (also see Jackson,
Xanthopoulos, \& Browne 2000) yield $\Rcusp=0.619\pm0.050$ in
V-band, $0.572\pm0.147$ in I-band, and $0.473\pm0.092$ in H-band.
The decline in $\Rcusp$ with wavelength suggests that there might
be some differential extinction between the images, but the
evidence is weak because the three measurements are formally
consistent within the errors.  The weighted average of the
optical/near-IR measurements is $0.585\pm0.042$.

The difference between the radio and optical results is very
interesting.  At radio wavelengths the overlap between the observed
and predicted probability curves in \reffig{joint} is small but
non-negligible: the curves overlap at 2.1--3.7\% depending on the
input data used in the Monte Carlo simulations.  Thus, there is
evidence for a radio flux ratio anomaly, but only at the 96--98\%
confidence level.  At optical/near-IR wavelengths, by contrast, there
is no overlap between the observed and predicted probability curves,
and hence evidence for an optical flux ratio anomaly at high confidence.
Both conclusions are unaffected by octopole perturbations in the
lens potential, and more generally by changes to the input data in
the Monte Carlo simulations.  The difference between the radio and
optical results could indicate that the optical flux ratio anomaly
is caused by a star (microlensing) or some other object with a
characteristic size smaller than a typical dark matter subhalo.
That possibility makes B0712+472 a promising system for optical
monitoring to look for variability that would indicate microlensing.

\subsubsection{1RXS~J1131$-$1231}

Sluse et al.\ (2003) present three measurements of the optical flux
ratios of 1RXS~J1131$-$1231.  Observations from 2 May 2002 yield
$\Rcusp=0.350\pm0.021$ in V-band, while observations from 18 December
2002 yield $\Rcusp=0.353\pm0.031$ in V-band and $0.367\pm0.031$ in
R-band.  It is interesting that the total flux of the system varied
by $0.29\pm0.04$~mag between May and December, yet the flux ratios
and $\Rcusp$ values are essentially identical.  The likely explanation
is that the source varied over the 7-month time scale, but the short
time delays between the bright images A, B, C (predicted to be
$<$1~day) kept the flux ratios essential constant.  The fact that
$\Rcusp$ is the same at different epochs and in different passbands
indicates that there are no large systematic uncertainties due to
the time delays or electromagnetic effects.

The weighted average of the $\Rcusp$ values is $0.355\pm0.015$.
\reffig{joint} shows that this value lies well outside the predicted
distributions, implying a strong flux ratio anomaly, and that this
conclusion is insensitive to changes in the input data in the Monte
Carlo simulations.

\subsubsection{RX~J0911+0551}

RX~J0911+0551 is the only cusp lens that shows significant evidence
for differential extinction between the images.  In the CASTLES data
(also see Burud et al.\ 1998), image A$_1$ has colors $V-H=1.24\pm0.04$
and $I-H=0.79\pm0.04$ while images A$_2$ and A$_3$ both have colors
$V-H=1.54\pm0.05$ and $I-H=1.01\pm0.04$.  Attributing the color
difference to dust in the lens galaxy, we estimate a differential
extinction between A$_1$ and the A$_2$/A$_3$ pair of $(0.44,0.32,0.12)$
mag in $(V,I,H)$ (following the analysis of Falco et al.\ 1999, using
a redshifted $R_V=3.1$ extinction curve from Cardelli, Clayton \&
Mathis 1989).\footnote{M\"oller et al.\ (2003) claim that the
extinction correction for RX~J0911+0551 is highly uncertain, but we
find that not to be the case.}  Correcting for the reddening changes
$\Rcusp$ from 0.226 for the raw H-band flux ratios to
$\Rcusp=0.192\pm0.011$ for the de-reddened data.

\reffig{joint} compares this value with the predictions of smooth
lens potentials.  Perhaps the most interesting point is that
RX~J0911+0551 is the only system where octopole perturbations affect
the judgment about a flux ratio anomaly.  The Monte Carlo simulations
that include octopole modes (based on the Bender et al.\ or Saglia
et al.\ input data) have a long tail to high $\Rcusp$ values that
is absent from simulations lacking such modes (based on the
J{\o}rgensen et al.\ input data, or re-running the Bender et al.\
or Saglia et al.\ data with the octopole terms removed [not shown]).
The octopole modes apparently allow smooth lens potentials to
create a small but significant number of image triplets with the
same angle $\theta$ and separation $d$ as RX~J0911+0551 that have
relatively high values of $\Rcusp$.  The reason we see such an effect
here, but not in the other lenses we study, is that RX~J0911+0551 is
the only system where the source lies near a cusp on the {\it minor\/}
axis of the galaxy.  As we showed in \refsec{simple}, octopole
perturbations mainly affect sources near a minor-axis cusp.  The
octopole modes cause the observed and predicted probability curves
in \reffig{joint} overlap at 1.7\% for the Bender et al.\ input
data or 8.6\% for the Saglia et al.\ input data.  We caution against
overinterpreting these numbers, however.  Because the Bender et
al.\ and Saglia et al.\ samples include just 87 and 54 ellipticity
and octopole measurements, respectively, we may worry about
small-number statistics and sample variance.  Indeed, the difference
between the $\Rcusp$ distributions predicted by these two samples
suggests that a robust analysis of the cusp relation for minor-axis
lenses like RX~J0911+0551 will require larger samples with better
determinations of the ellipticity and multipole distributions.

\begin{figure}[t]
\centerline{\epsfxsize=3.0in \epsfbox{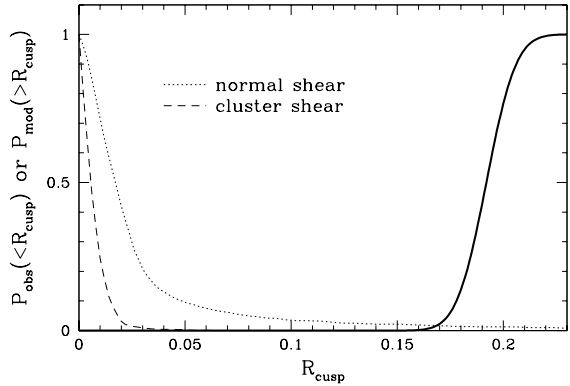}}
\caption{
Results for RX~J0911+0551 for the ``normal shear'' and ``cluster
shear'' cases, using the Bender et al.\ input data.
}\label{fig:0911clus}
\end{figure}

The analysis of RX~J0911+0551 is actually even more complicated,
because the lens galaxy lies in a cluster environment that
contributes a large shear $\gamma \simeq 0.3$ to the lens potential
(Kneib, Cohen, \& Hjorth 2000).  Since this shear is almost $4\sigma$
above the median of our assumed distribution, it is not likely to
be well represented in our fiducial Monte Carlo simulations.  To
examine the effects of such a large shear, we use a new Monte Carlo
simulation with the shear amplitude fixed at $\gamma=0.3$.  The
orientation of the shear is still random, and we adopt the Bender
et al.\ (1989) input data for the ellipticity and octopole.
\reffig{0911clus} shows that lens potentials with a large shear
produce a {\it narrower\/} range of $\Rcusp$ values than lens
potentials with the fiducial shear distribution.  The likely
explanation is that an octopole term has the most effect when
aligned with the quadrupole moment of the potential, which cannot
happen when the quadrupole is dominated by a large, randomly
oriented shear.  As a result, the $\Rcusp$ distribution predicted
when the shear is large no longer overlaps with the observed value
of $\Rcusp$.

We believe that the ``cluster shear'' case provides strong evidence
for a flux ratio anomaly in RX~J0911+0551.  However, it will be
important to examine more carefully whether multipole modes can
compromise this conclusion.  The most straightforward approach will
be direct lens modeling.  It will definitely be possible to fit the
current data exactly by allowing enough multipole modes in the models
(see Evans \& Witt 2002 for examples).  The crucial test would be to
obtain deep near-IR imaging to try to find an Einstein ring image
of the quasar host galaxy (e.g., Impey et al.\ 1998; Keeton et al.\
2000; Kochanek et al.\ 2001).  The additional constraints from the
Einstein ring would greatly restrict the space of allowed models
and determine whether the multipole modes required by the flux
ratios are acceptable or not (see Kochanek \& Dalal 2003).

If the putative flux ratio anomaly in RX~J0911+0551 is confirmed,
its interpretation may be ambiguous.  With only broad-band optical
and near-IR flux ratios it is impossible to determine whether the
small-scale structure implied by the anomaly corresponds to
micro-lensing or milli-lensing --- or to one or more dwarf galaxies,
which tend to be abundant in cluster environments (Trentham 1997,
Smith, Driver, \& Phillipps 1997; Driver, Couch, \& Phillipps 1998;
Zabludoff \& Mulchaey 2000).  In fact, the CASTLES images do reveal
a faint satellite galaxy lying 1\farcs0 away from the main lens
galaxy (although not close to any of the lensed images).  Analysis
of lens models that explicitly include the satellite suggest that
it cannot explain the observed flux ratios; as seen in \reftab{data},
such models still predict $\Rcusp<0.01$ for the cusp triplet.  Still,
we cannot presently rule out the possibility that the anomaly is
caused by an as-yet undetected but otherwise unremarkable dwarf
galaxy.  Perhaps the best prospect for determining whether the
anomaly is caused by a star, a dark matter clump, or a dwarf galaxy
would be high-resolution resolved spectroscopy of the three images
(see Moustakas \& Metcalf 2003).

\subsubsection{B1422+231}

Patnaik \& Narasimha (2001) report 61 measurements of the radio
fluxes for B1422+231 at 8.4 and 15~GHz.  The mean and scatter for
$\Rcusp$ from all of the measurements is $0.179\pm0.006$.  The error
on $\Rcusp$ obtained directly from the individual flux errors is
comparable to or smaller than the scatter.  Koopmans et al.\ (2003a)
report 41 measurements at 5~GHz that yield $\Rcusp=0.187\pm0.004$,
in reasonable agreement with the value from the Patnaik \& Narasimha
(2001) data.

The optical and near-IR data from CASTLES yield $\Rcusp=0.223\pm0.055$
in V-band, $0.222\pm0.038$ in I-band, and $0.175\pm0.015$ in H-band.
The change from V/I to H is comparable to the uncertainties.  The
weighted average of the optical/near-IR results is
$\Rcusp=0.184\pm0.014$, in remarkably good agreement with the radio
results.

\reffigs{Rcusp-obs}{joint} show that B1422+231 is not an outlier
in the cusp relation according to our analysis.  This result is
interesting in light of detailed modeling that shows B1422+231 to
have a flux ratio anomaly (Mao \& Schneider 1998; Brada\v{c} et
al.\ 2002; Metcalf \& Zhao 2002).  The reason for the difference
is that we are focusing on the {\it maximum allowed\/} values of
$\Rcusp$ for given values of $\theta$ and $d$.  We do not consider
whether the potentials that can produce those values are at all
consistent with the rest of the lens data.  In other words, we are
throwing away reliable information in our attempt to be as general
and robust as possible.  Metcalf \& Zhao (2002) demonstrate that
imposing the position constraints via detailed lens modeling
provides a more sophisticated analysis that has more power to
identify flux ratio anomalies.  But because it relies on specific
modeling, we believe that such model fitting is less robust and
generic than the cusp relation analysis.

\subsubsection{SDSS~J0924+0219}

The lens SDSS~J0924+0219 is not a typical cusp configuration but
rather more like a cross configuration.  Nevertheless, it does have
what seems to be an enormous flux ratio anomaly.  The anomaly appears
as such a strong suppression of one of the negative-parity images
that the system was initially thought to have just three images, and
the fourth image (D) was identified only after subtraction of the
three main images and the lens galaxy (Inada et al.\ 2003).  Because
image D is a factor of 14 fainter than image A, the triplet ABD 
(with $\theta=156\arcdeg$ and $d/\Rein=2.1$) has $\Rcusp$ values
of $0.916\pm0.009$ in $g$-band, $0.903\pm0.005$ in $r$-band, and
$0.894\pm0.005$ in $i$-band.  The good agreement between the
different passbands suggests that there is little differential
extinction between the images.  The weighted average of the
measurements is $\Rcusp=0.901\pm0.003$.

The fact that SDSS~J0924+0219 is not a tight cusp configuration
means that smooth potentials can produce a fairly broad distribution
of $\Rcusp$ values (see \reffig{joint}).  Nevertheless, the observed
value of $\Rcusp$ is so large that it still lies well outside the
predicted distribution.  In other words, the anomaly in SDSS~J0924+0219
is so strong that it is identified by the cusp relation analysis even
though the lens is not a proper cusp configuration.  This conclusion
is insensitive to changes in the input data of the Monte Carlo
simulations, and is therefore robust provided that the lens galaxy
is early-type.  Because the smallest image separation is still
relatively large ($d/\Rein=2$, $\theta=156\arcdeg$), a spiral galaxy
might in principle have structure on scales appropriate to explain
the anomaly.  However, the strength of the anomaly would require
significant structure, which would probably affect not just image
D but all four images; and there is no current evidence that the
lens is a spiral (Inada et al.\ 2003).  In any case, the flux ratio
anomaly in SDSS~J0924+0219 indicates that there is some strong and
interesting structure in the lens potential that is generally
incompatible with the known properties of the luminous components
of early-type galaxies.

\subsection{Comments}

To conclude this section, we remark that it is surprising to see how
large $\Rcusp$ can be for realistic lenses, {\it even in the absence
of small-scale structure}.  Image triplets must be quite tight
(small $\theta$ and $d$) in order to be guaranteed of satisfying
the cusp relation reasonably well.  For example, to satisfy the
relation with $\Rcusp<0.1$ at 99\% confidence, the opening angle
must be $\theta \lesssim 30\arcdeg$.  The parameter space that gives
rise to such image configurations is quite small; in our Monte Carlo
simulations, only $\sim$0.1\% of 4-image lenses have cusp triplets
that are this tight.  While this estimate omits magnification bias
(which will favor tight configurations), it does suggest that cusp
image configurations that are close to ideal are likely to be rare.
The cusp relation can still be a valuable tool for identifying flux
ratio anomalies, but it must be used with some care.

We have identified flux ratio anomalies that cannot satisfy the
cusp relation for any (reasonable) combination of shear,
ellipticity, and octopole modes.  While we do not believe that
significant high-order multipole perturbations are common in real
early-type galaxies, it is nevertheless interesting to consider what
modes would be required to explain the anomalies.  Using the analytic
results from \refsec{simple} (\reffig{anal2} in particular), we
estimate that B2045+265 would require 1.5\% density fluctuations in
a mode with $m \gtrsim 16$.  The B0712+472 optical anomaly would
require 4.5\% fluctuations with $m \gtrsim 8$.  1RXS~J1131$-$1231
would require 1.5\% fluctuations with $m \gtrsim 10$.  
SDSS~J0924+0219, being a cross rather than a cusp configuration,
does not require a particularly high-order mode ($m \gtrsim 4$),
but it would require a perturbation amplitude of at least 12\%.
The estimated amplitudes are based on an assumed ellipticity $e=0.3$,
and they would increase or decrease as the ellipticity is increased
or decreased.  Note that these estimates are consistent with our
rule of thumb from \refsec{simple} that a large $\Rcusp$ value in
a triplet with angle $\theta$ requires multipole modes with
$m \gtrsim 640\arcdeg/\theta$.  Thus, the most general statement
about the lens potential that can be derived from a flux ratio
anomaly is that there must be significant structure on the scale
of the separation between the images.

\section{Conclusions}
\label{sec:concl}

The images associated with an ideal cusp catastrophe satisfy
universal position and magnification relations, the most interesting
of which says that the signed magnifications of the three images
should sum to zero.  (The other relations have less practical value
because they involve unobservable quantities.)  A violation of this
relation indicates that the catastrophe is not an ideal cusp ---
that the lens potential contains terms beyond fourth order (third
order in the lens equation; see eq.~\ref{eq:le-cusp0}).  A
significant violation of the cusp relation can imply that an
observed 4-image lens has significant structure in the lens potential
on scales comparable to or smaller than the separation between the
images.

An important caveat is that the ideal cusp relation is not expected
to hold exactly for real lenses, because real lens galaxies are not
ideal cusps.  It is therefore crucial to understand the degree to
which simple aspects of real lens potentials (the radial density
profile, ellipticity, multipole density fluctuations, and tidal
shear) can cause deviations from the ideal cusp relation.  After
showing that the radial profile has little effect on the analysis,
we have adopted observationally-motivated distributions for the
shear, ellipticity, and octopole perturbations and derived the
resulting distribution for the quantity
$\Rcusp = (|F_1+F_2+F_3|)/(|F_1|+|F_2|+|F_3|)$, where the $F_i$
are the signed fluxes for a triplet of images.  The distribution
describes the range of $\Rcusp$ values that can be produced by lens
potentials containing realistic shears, ellipticities, and octopole
modes.  If an observed 4-image lens has an $\Rcusp$ value lying
outside this range, we can conclude that its lens potential must
have some kind of structure on scales comparable to or smaller
than the image separation that is not represented in the luminous
properties of early-type galaxies or the external shears expected
from their typical environments.  This structure might be large,
low-order multipole moments, significant power in moments higher
than octopole order, or some kind of compact structure.  Analysis
of the cusp relation therefore provides a new way to search for
cases of milli-lensing, micro-lensing, or other interesting
small-scale structure, which is an attractive alternative to
global lens modeling because it relies on a local analysis of
lensing near a cusp to probe localized structure in the lens
potential.

When we examine the nineteen known 4-image lenses, we find
evidence for flux ratio anomalies (violations of the cusp relation)
in four of the five lenses with cusp image configurations, plus
one other lens.
First, B2045+265 has a very strong flux ratio anomaly at both radio
and optical/near-IR wavelengths, which if attributed to multipole
density fluctuations would require significant structure in modes
with $m \gtrsim 16$.
Second, B0712+472 has an anomaly that is very strong at
optical/near-IR wavelengths but marginal at radio wavelengths,
which would require structure in multipole modes with $m \gtrsim 8$.
The fact that the anomaly is much stronger at optical/near-IR
wavelengths than at radio wavelengths might suggest that it is due
to micro-lensing rather than milli-lensing, but further study is
required to test that hypothesis.
Third, 1RXS~J1131$-$1231 has a strong anomaly at optical wavelengths,
which would require structure in multipole modes with $m \gtrsim 10$.
Fourth, RX~J0911+0551 has a value of $\Rcusp$ that suggests a flux
ratio anomaly but whose interpretation is somewhat complicated.
As the only known system where the source lies near a cusp on the
minor (rather than major) axis of the lens potential, RX~J0911+0551
is the only one where octopole modes can allow simple lens potentials
to produce relatively high values of $\Rcusp$.  Strong conclusions
about the putative anomaly will require better knowledge of the
distribution of octopole amplitudes in real galaxies.  Still, we
believe that the evidence for a flux ratio anomaly in this system
is good, especially when the fact that the lens lies in a cluster
environment is taken into account.
Finally, SDSS~J0924+0219 is an intriguing system with a flux ratio
anomaly so strong that it is identified by the cusp relation analysis
even though the lens does not have a cusp image configuration.

Interestingly, we find that the cusp lens B1422+231 does not
obviously violate the cusp relation, even though it is known to
have a flux ratio anomaly from detailed lens modeling (Mao \&
Schneider 1998; Brada\v{c} et al.\ 2002; Metcalf \& Zhao 2002).
This illustrates an important point: in analyzing the cusp relation
we have ignored constraints from the observed image positions,
in order to be general and to avoid explicit modeling as much as
possible.  The idea is that an analysis based purely on magnification
relations is the most robust and conservative way to identify flux
ratio anomalies.  Adding constraints from the image positions
yields an analysis that is more sophisticated and has more power to
identify flux ratio anomalies (see Metcalf \& Zhao 2002 for a good
example), but it requires detailed modeling and is therefore less
generic.  Furthermore, even when direct modeling suggests that a
set of flux ratios cannot be fit by simple lens potentials, it may
not reveal {\it why\/} that is the case.  The cusp relation
immediately pinpoints the cause of the failure.  For these reasons,
we believe that the cusp relation analysis is the best place to
start when seeking to identify lenses with flux ratio anomalies
(at least for lenses with cusp configurations).  If an observed
lens violates the cusp relation then the anomaly is unambiguous
and easy to understand.  If it does not violate the cusp relation,
it may still have an anomaly but may only be revealed by more
sophisticated and less model-independent analyses.

Although we have argued that flux ratio anomalies exist and
indicate a strong need for small-scale structure in lens galaxies,
we cannot draw strong conclusions about what that structure must
be.  Analyses based on image positions and broad-band flux
ratios can only give an upper limit on the characteristic angular
scale of the structure implies by flux ratio anomalies (also see
Evans \& Witt 2002; Quadri et al.\ 2003; M\"oller et al.\ 2003).
Plausibility arguments and astrophysical expectations might be
invoked to favor one possibility over another.  For example,
stars are known and dark matter clumps expected to be abundant
in lens galaxies, while high-order multipole modes are not, so
one might prefer to attribute flux ratio anomalies to stars and
dark matter clumps.  But it is important to acknowledge the
prejudices inherent in such arguments.

Fortunately, there are excellent prospects for obtaining data that
move beyond broad-band fluxes to determine the nature of the
small-scale structure.  One possibility is to look for time
variability that is an unmistakable signature of microlensing
by stars (e.g., Wo\'zniak et al.\ 2000; Schechter et al.\ 2003).
A second good possibility is to look for Einstein ring images of
the quasar host galaxy in deep near-IR images.  The Einstein rings
would vastly improve the constraints on the global properties of
the lens potential and reveal whether high-order multipole modes
are acceptable (Kochanek et al.\ 2001; Kochanek \& Dalal 2003).
A third possibility is to use an aspect of the cusp relation that
we have ignored, namely the sign of $\Rcusp$.  Several recent
studies have indicated that localized structure is sensitive to
the image parities, most often suppressing negative-parity images
and occasionally increasing the magnification of positive-parity
images (Schechter \& Wambsganss 2002; Keeton 2003), while global
modes make no distinction between different images.  Thus, in an
ensemble of lenses with flux ratio anomalies, the presence or
absence of skewness in the set of $\Rcusp$ values could reveal
whether the small-scale structure is local or global.  Indeed,
Kochanek \& Dalal (2003) already find evidence that many flux
ratio anomalies can be attributed to suppression of negative-parity
images, which suggests that many flux ratio anomalies are in fact
caused by milli-lensing and/or micro-lensing.  Finally, resolved
high-resolution spectroscopy of systems with flux ratio anomalies
offers the intriguing possibility of determining the physical scale
of the structure (Moustakas \& Metcalf 2003; Wisotzki et al.\ 2003).
Thus, the future is very bright for using detailed study of lenses
with flux ratio anomalies to learn about small-scale structure in
lens galaxies.  Since all of these applications begin with the
identification of flux ratio anomalies, the cusp relation analysis
will be of fundamental importance to all of them.

\acknowledgements

We would like to thank the organizers of the 2002 Ringberg Workshop
on Gravitational Lenses, where this work was conceived, and Paul
Schechter for his list of 13 Provocations; this work addresses
Provocation \#9.  We also thank Rennan Barkana, Neal Dalal, Gil
Holder, and Chris Kochanek for interesting discussions.  We thank
Christina Turner for extensive work in assembling data and running
models for the known lenses.  Finally, we thank the anonymous
referee for an insightful critique that helped us substantially
improve the paper.
This work was supported by NASA through Hubble Fellowship grants
from the Space Telescope Science Institute, which is operated by
the Association of Universities for Research in Astronomy, Inc.,
under NASA contract NAS5-26555; by JPL contract 1226901; by an
Alfred P.\ Sloan Research Fellowship; and by NSF Career grant
DMS-98-96274.

\appendix

\section{Universal Relations for Cusps}
\label{app:analytics}

In this Appendix we study the general properties of the lensing map
near a cusp catastrophe to derive generic relations between the
image positions and magnifications that should be satisfied whenever
the source is sufficiently close to a cusp.  This analysis applies
to ordinary cusps, in which the two branches of the curve approach
the cusp from opposite sides of the line that is tangent to the
cusp.  The analysis may not be valid for ramphoid cusps, an
alternate situation where the two branches approach the cusp from
the same side of the tangent line (for examples, see Petters \&
Wicklin 1995; Oguri et al.\ 2003).  Only ordinary cusps have been
observed, and ramphoid cusps are expected to be rare in lensing
situations of astrophysical interest.

\subsection{Local orthogonal coordinates}
\label{app:orthog}

Consider the lens equation
\begin{equation}
  \bby = \bx - \grad \psi (\bx) .
\end{equation}
Assume that the induced lensing map,
$\bme (\bx) = \bx - \grad \psi (\bx)$, from the lens plane $L$ to
the light source plane $S$ is locally stable, which yields that the
caustics of $\bme$ are either folds or cusps (Petters et al.\ 2001,
p.~294). Then translate coordinates in the lens and light source
planes so that the cusp point of interest is at the origin and the
critical point mapping to the cusp is also at the origin, i.e.,
$\bme (\bmo) = \bmo$. By abuse of notation, the resulting
translations of $\bx$ and $\bby$ will still be denoted by those
symbols.

We now define a change of coordinates about the origins of the lens and light
source planes: $\bx \rightarrow \bt$, $\bby \rightarrow \bu$. The
coefficients of the quadratic terms of the Taylor expansion of
$\psi$ at the origin are
\begin{equation} \label{eq:ahat}
  \ha = \frac{1}{2} \psi_{11} (\bmo), \qquad
  \hb =  \psi_{12} (\bmo), \qquad
  \hc =  \frac{1}{2} \psi_{22} (\bmo),
\end{equation}
where the subscripts indicate partial derivatives relative to
$\bx = (x_1, x_2)$. Since the origin is a cusp critical point,
$(1-2\ha)$ and $(1-2\hc)$ cannot both vanish (Petters et al.\ 2001,
p.~349). Without loss of generality, we shall assume that
$(1-2\ha) \neq 0$. Define an orthogonal matrix as follows:
\begin{equation}
  {\rm M} = \frac{1}{\sqrt{(1 - 2\ha)^2 + \hb^2}} \:
    \left[ \begin{array}{cc}
      1 - 2\ha & -\hb \\
      \hb & 1 - 2\ha
    \end{array} \right].
\end{equation}
The new coordinate systems are defined by
\begin{equation} \label{eq:coord-change}
  \bt = (\theta_1, \theta_2) \equiv  {\rm M}\, \bx \, ,
  \qquad
  \bu = (u_1, u_2) \equiv {\rm M}\, \bby \, .
\end{equation}
Note that this is the {\it same\/} coordinate change in the lens
and light source planes, and the coordinates depend on the
potential.

It can be proven rigorously that the lensing map $\bme$ in the
orthogonal coordinates (\ref{eq:coord-change}) can be approximated
in a neighborhood of the cusp critical point at the origin by a
simple polynomial mapping (Petters et al.\ 2001, pp.~341-353; also
see Schneider et al.\ 1992, p.~193):
\begin{equation} \label{eq:le-cusp}
  u_1 = c \ \theta_1 \  + \  \frac{b}{2}\  \theta^2_2 \, , \qquad
  u_2 = b \  \theta_1 \  \theta_2 + a \  \theta_2^3 \, ,
\end{equation}
where
\begin{eqnarray}
  a &=& -\frac{1}{6} \psi_{2222}(\bmo) \, ,
  \qquad b = -\psi_{122} (\bmo) \neq 0 \, , \nonumber\\
  c &=& 1 - \psi_{11} (\bmo) \neq 0 \, ,
  \qquad  2ac - b^2 \neq 0 \, .
  \label{eqn:coeff}
\end{eqnarray}
The partial derivatives of $\psi$ are relative to the original
coordinates $\bx = (x_1, x_2)$. The origin in the light source
plane is called a {\it positive cusp\/} if $2ac > b^2 $ and a
{\it negative cusp\/} if $2ac < b^2$. A source inside a positive
cusp has, locally, two images with positive parity and one with
negative parity; the reverse is true for negative cusps.

\subsection{Position relations}
\label{app:posreln}

Using the lens equation (\ref{eq:le-cusp}), the three local lensed
images associated with a source inside and close to the cusp have
the following positions (e.g., Gaudi \& Petters 2002):
\begin{equation} \label{eq:image-positions}
  \bt_i = \left(\frac{u_1}{c}  - \frac{b}{2c} z_i^2 \ , \ z_i \right),
  \qquad i=1,2,3 \,.
\end{equation}
The $z_i$ are the three real solutions of the cubic equation
\begin{equation} \label{eq:cubicpol}
  z^3 \ + \  \ttp \ z \ + \  \ttq  =  0 \, ,
\end{equation}
where
\begin{equation} \label{eqn:defpq}
  \ttp = \frac{2 b }{2ac - b^2}\, u_1 \equiv \hp \ u_1\, ,
  \qquad \ttq = -\frac{2c}{2ac - b^2}\, u_2 \equiv - \hq \ u_2\, .
\end{equation}
Note that when the source is inside the cusp the discriminant,
\begin{equation} \label{eq:discr}
  D = \left(\frac{\ttp}{3}\right)^3 + \left(\frac{\ttq}{2}\right)^2
  = \frac{4 (\hp \, u_1)^3 + 27 (\hq \, u_2)^2}{108}\ ,
\end{equation}
is negative so \eq{cubicpol} does have three real roots.

The usual factoring of a cubic polynomial yields:
\begin{eqnarray}
  0 &=& (z-z_1)(z-z_2)(z-z_3) \, , \\
  &=& z^3 \ - \ [z_1 + z_2 + z_3]\,z^2
    \ + \ [z_1 z_2 + z_1 z_3 + z_2 z_3]\,z \ - \ [z_1 z_2 z_3]\, .
\end{eqnarray}
Identifying coefficients with \eq{cubicpol} yields three relations
between the image positions:
\begin{eqnarray}
z_1 \ +\ z_2\  + \ z_3 &=& 0 \, ,
\label{eq:univ-position-1} \\
z_1 z_2 \ + \  z_1 z_3 \ + \  z_2 z_3 &=& \hp \, u_1 \, ,
\label{eq:univ-position-2} \\
z_1 z_2 z_3 &=& \hq \, u_2 \, .
\label{eq:univ-position-3}
\end{eqnarray}
These are universal relations satisfied by the image positions of
a triplet associated with a source near a cusp.  Two additional
relations can be obtained respectively by squaring
(\ref{eq:univ-position-1}) and using (\ref{eq:univ-position-2}),
and squaring (\ref{eq:univ-position-2}) and using
(\ref{eq:univ-position-1}):
\begin{eqnarray}
z^2_1 \ + \ z^2_2 \ + \ z^2_3 &=& - 2 \, \hp \, u_1 \, ,
\label{eq:univ-position-4} \\
(z_1 z_2)^2 \ + \ (z_1 z_3)^2 \ + \ (z_2 z_3)^2 &=& (\, \hp \, u_1 \,)^2 \, .
\label{eq:univ-position-5}
\end{eqnarray}
These relations are not independent of
(\ref{eq:univ-position-1})--(\ref{eq:univ-position-3}), but they
are more useful in certain circumstances (as seen below).

\subsection{Magnification relations}
\label{app:magreln}

The signed magnification of each image $\bt_i$ in the triplet
associated with the cusp is given by
\begin{equation} \label{eq:cusp-mag}
  \mu_i = \frac{1}{\det[\jac \bu](\bt_i)}
  = \frac{\hp}{b\,(\hp \, u_1 \ + \ 3 \, z_i^2)}\ , \qquad i=1,2,3 \, ,
\end{equation}
where $\jac \bu$ is the Jacobian matrix of the lensing map
(\ref{eq:le-cusp}). Note that $\jac \bu = {\rm M}\, \jac \bby$, so
with ${\rm M}$ an orthogonal matrix we verify that the
magnification is independent of our choice of coordinates:
$\det[\jac \bu] = \det[\jac \bby]$.

Three known universal relations between the magnifications $\mu_i$
are as follows (Schneider \& Weiss 1992; Zakharov 1995; Petters et
al.\ 2001, p.~339):
\begin{eqnarray}
\mu_1 \ + \ \mu_2\ + \ \mu_3 &=& 0 \, ,
\label{eq:univ-mag-1} \\
\mu_1 \mu_2 \ + \ \mu_1 \mu_3 \ + \ \mu_2 \mu_3  &=&
  -\, \frac{\hp^3 \, u_1}{36\, b^2 D}\ ,
\label{eq:univ-mag-2} \\
\mu_1 \mu_2 \mu_3 &=&  \frac{\hp^3}{108\, b^3 D}\ ,
\label{eq:univ-mag-3}
\end{eqnarray}
where $D$ is given by \eq{discr}. These relations can be verified
by direct calculation from (\ref{eq:cusp-mag}), using the position
relations (\ref{eq:univ-position-3})--(\ref{eq:univ-position-5})
for simplifications.
In analogy with the position relations, we can derive additional
magnification relations:
\begin{eqnarray}
\mu_1^2 \ + \ \mu_2^2 \ + \ \mu_3^2 &=& \frac{\hp^3 \, u_1}{18\, b^2 D}\ ,
\label{eq:univ-mag-4} \\
(\mu_1 \mu_2)^2 \ + \ (\mu_1 \mu_3)^2 \ + \ (\mu_2 \mu_3)^2 &=&
  \left(\frac{\hp^3 \, u_1}{36\, b^2 D}\right)^2 .
\label{eq:univ-mag-5}
\end{eqnarray}
These quadratic magnification sum rules have not appeared in the
literature before.

\section{Simple Lens Potentials}
\label{app:exact}

In this Appendix we derive exact solutions to the lens equation to
use as a benchmark for understanding the cusp relations.  Exact
solutions are possible only for certain lens potentials, and then
only for sources on a symmetry axis.  We consider two families of
potentials: a spherical galaxy with a power law density profile
plus an external shear; and a singular isothermal ellipsoid with
multipole density perturbations plus an external shear aligned with
the major or minor axis of the galaxy.

\subsection{Power law galaxy with shear}
\label{app:PL}

Consider the lens potential
\begin{equation}
  \psi(r,\phi) = \frac{1}{\alpha} \Rein^{2-\alpha} r^{\alpha}
    - \frac{\gamma}{2} r^2 \cos2\phi .
\end{equation}
The first term represents a spherical galaxy with a power law
profile for the surface mass density,
\begin{equation}
  \kappa(r) = \frac{ \Sigma(r) }{ \Sigma_{\rm crit} }
  = \frac{\alpha}{2} \left(\frac{\Rein}{r}\right)^{2-\alpha} ,
\end{equation}
where $\Rein$ is the Einstein radius. The case $\alpha=1$
corresponds to a singular isothermal sphere (SIS), while the cases
$\alpha<1$ and $\alpha>1$ correspond respectively to steeper and
shallower profiles, respectively. The second term in the potential
represents an external tidal shear with amplitude $\gamma$. Without
loss of generality, we are working in coordinates such that the
shear is aligned with the horizontal axis ($\gamma>0$) or the
vertical axis ($\gamma<0$).

Using polar coordinates in the image plane and Cartesian
coordinates in the source plane, the lens equation has the form
\begin{eqnarray}
  y_1 &=& r\cos\phi \left[ 1 + \gamma - \left(\frac{\Rein}{r}\right)^{2-\alpha}
    \right], \label{eq:PLeq1} \\
  y_2 &=& r\sin\phi \left[ 1 - \gamma - \left(\frac{\Rein}{r}\right)^{2-\alpha}
    \right], \label{eq:PLeq2}
\end{eqnarray}
and the lensing magnification $\mu$ is given by
\begin{equation}
  \mu^{-1} = 1-\gamma^2 - (1-\alpha)\left(\frac{\Rein}{r}\right)^{4-2\alpha}
    - \left(\frac{\Rein}{r}\right)^{2-\alpha} \left[ \alpha +
    (2-\alpha)\gamma\cos2\phi \right] .  \label{eq:PLminv}
\end{equation}
The critical curve in the image plane is the curve where
$\mu^{-1}=0$, and it maps to the caustic in the source plane. The
caustic has a cusp on the horizontal axis at position $(y_{1c},0)$,
which corresponds to a point on the critical curve at position
$(x_{1c},0)$, where
\begin{eqnarray}
  x_{1c} &=& \frac{       \Rein}{ (1-\gamma)^{1/(2-\alpha)} }\ ,\label{eq:PLx1c}
\\
  y_{1c} &=& \frac{2\gamma\Rein}{ (1-\gamma)^{1/(2-\alpha)} }\ .\label{eq:PLy1c}
\end{eqnarray}

The Taylor series coefficients used to define the local orthogonal
coordinate system in \refapp{orthog} are as follows:
\begin{eqnarray}
  \ha &=& \frac{1}{2}\,\left[-1+\alpha(1-\gamma)\right]\, ,
  \qquad
  \hb = 0\, ,
  \qquad
  \hc = \frac{1}{2}\ , \\
  a &=& \frac{2-\alpha}{2\Rein^2}\, (1-\gamma)^{(4-\alpha)/(2-\alpha)}\, ,
  \qquad
  c  =  2 - \alpha(1-\gamma)\, , \\
  b &=& \frac{2-\alpha}{\Rein}\, (1-\gamma)^{(3-\alpha)/(2-\alpha)}\, ,
  \qquad
  \hp = \frac{ \Rein }{ \gamma (1-\gamma)^{1/(2-\alpha)} }\ .
\end{eqnarray}
Hence the transformation matrix ${\rm M}$ in \eq{coord-change} is
the identity matrix, so the $\bt$ and $\bu$ coordinate systems are
simply the $\bx$ and $\bby$ coordinate systems translated so the
cusp point is at the origin.

Note that although the potential $\psi$ is not well defined in the
limit $\alpha \to 0$, the lens equation and magnification and other
quantities are perfectly well defined and correspond to a point
mass in a shear field. Furthermore, in this limit the surface mass
density $\Sigma$ is a $\delta$-function as expected for a point
mass. Hence in this formalism we can consider the case $\alpha=0$
to correspond to a point mass lens.

Consider a source on the horizontal axis inside the caustic; for
$\gamma>0$ ($\gamma<0$) this correspond to the major (minor) axis
of the lens potential. The lens equation can be solved exactly
because of symmetry.  There is at least one image on the
$x_1$-axis,\footnote{There may or may not be an image on the
$x_1$-axis on the opposite side of the origin from the source,
depending on whether the cusp is ``clothed'' or ``naked'' (e.g.,
Schneider et al.\ 1992; Petters et al.\ 2001).  We are interested
only in the image on the $x_1$-axis on the same side of the origin
as the source.} and two images off the $x_1$-axis.  By symmetry,
the two off-axis images are identical modulo some signs.  To find
the positions of these two images, note that with $y_2 = 0$ and
$\phi \ne 0$ the only way for \eq{PLeq2} to be satisfied is for the
term in square brackets to vanish.  This condition yields the polar
radius, which can then be substituted into \eq{PLeq1} to find the
polar angle.  Thus, the positions of the two off-axis images, which
we label A and C, are
\begin{equation}
  r_A = r_C = \frac{ \Rein }{ (1-\gamma)^{1/(2-\alpha)} }\ ,
\end{equation}
\begin{equation}
  \phi_A = -\phi_C = \cos^{-1}\left(\frac{ y_1 }{ y_{1c} }\right) ,
\end{equation}
where $y_{1c}$ is given by eq.~(\ref{eq:PLy1c}). Their
magnifications of these two images are
\begin{equation}
  \mu_A = \mu_C = \left\{ 2 \gamma (1-\gamma) (2-\alpha)
    \left[ 1 - \left(\frac{ y_1 }{ y_{1c} }\right)^2 \right] \right\}^{-1} .
\end{equation}
The image separation for this triplet is simply $d = 2 r_A \sin\phi_A$.

For the on-axis image, which we label B, eq.~(\ref{eq:PLeq2}) is
satisfied trivially ($y_2 = 0$ and $\phi_B = 0$).
Eq.~(\ref{eq:PLeq1}) can be solved analytically for integer
and half-integer values of $\alpha$, yielding:
\begin{eqnarray}
\alpha=0:&&\quad
  r_B = \frac{ y_1 + \sqrt{y_1^2 + 4(1+\gamma)\Rein^2} }{ 2(1+\gamma) } \\
\alpha=\frac{1}{2}:&&\quad
  r_B = \frac{ (\xi+y_1)^2 }{ 3\xi(1+\gamma) } \qquad\mbox{(see below)}\\
\alpha=1:&&\quad
  r_B = \frac{ y_1 + \Rein }{ 1+\gamma } \\
\alpha=\frac{3}{2}:&&\quad
  r_B = \frac{ \Rein + 2(1+\gamma)y_1 + \sqrt{\Rein [\Rein+4(1+\gamma)y_1] } }
    { 2(1+\gamma)^2 }
\end{eqnarray}
In the result for $\alpha=1/2$, $\xi$ is given by
\begin{equation}
  \xi^3 = \frac{27}{2}(1+\gamma)\Rein^3 - y_1^3
    + \frac{3}{2} \left\{ 3(1+\gamma) \Rein^3 \left[ 27(1+\gamma)\Rein^3
    - 4 y_1^3 \right] \right\}^{1/2} .
\end{equation}
The magnification $\mu_B$ of image B can then be computed from
\eq{PLminv}.

This analysis applies only to sources on the symmetry axis of the
lens, but we can begin to understand what happens when the source
is moved off-axis by examining derivatives with respect to $y_2$.
The first derivative of $\Rcusp$ vanishes by symmetry,
\begin{equation}
  \frac{\partial\Rcusp}{\partial y_2}\biggr|_{y_2=0} = 0\, ,
\end{equation}
so the axis is a local extremum.  The second derivative, which
determines whether it is a local maximum or minimum, can be computed
explicitly for an SIS plus shear potential. After lengthy but
straightforward algebra, we find
\begin{eqnarray}
  \frac{\partial^2\Rcusp}{\partial y_2^2}\biggr|_{y_2=0}
  &=& -\ \frac{ (1+\gamma)^2 }{ \gamma^2 \sin^4(\theta/2)
    [3+2\gamma+(1+\gamma)\cos(\theta/2)+\gamma\cos\theta]^2 } \nonumber\\
  && \times \biggl[ 4\gamma(1-\gamma)
     + (7+6\gamma-\gamma^2)\cos(\theta/2)
     + 16\gamma(2+\gamma)\cos^2(\theta/2) \nonumber\\
  && \qquad
     + (5+18\gamma+13\gamma^2)\cos^3(\theta/2)
     + 12\gamma(1+3\gamma)\cos^4(\theta/2) \biggr] .
\end{eqnarray}
The factor on the first line is manifestly negative, while the
quantity in square brackets on the second and third lines is
positive over the entire interesting range $0<\theta<\pi$ and
$|\gamma|<1$.  Thus, the second derivative is negative, and hence
$\Rcusp$ is a maximum on the axis.  While this proof formally holds
only for the SIS plus shear potential, intuition and Monte Carlo
simulations suggest that it is not restricted to this model.  On-axis
sources therefore provide a simple and important upper bound on
$\Rcusp$.

\subsection{Generalized ``isothermal'' galaxy with shear}
\label{app:GISO}

Consider the potential/density pair
\begin{eqnarray}
  \psi(r,\phi) &=& r F(\phi) - \frac{\gamma}{2} r^2 \cos2\phi\,, \\
  \kappa(r,\phi) &=& \frac{G(\phi)}{2r}\,,
\end{eqnarray}
where, from the Poisson equation, $F(\phi)$ and $G(\phi)$ are related
by
\begin{equation}
  G(\phi) = F(\phi) + F''(\phi)\,.
\end{equation}
The density and the first term in the potential correspond to a
mass distribution that is scale-free in the radial direction and
produces a flat rotation curve; such a model is often referred to
as ``isothermal'' in the lensing literature. The mass distribution
is allowed to have an arbitrary angular shape specified by the
functions $F(\phi)$ and $G(\phi)$. This family of models
includes both the singular isothermal ellipsoid and the singular
isothermal elliptical potential but is much more general, and its
lensing properties have been studied by Witt, Mao, \& Keeton (2000),
Evans \& Witt (2001, 2002), and Zhao \& Pronk (2001).  The second
term represents an external tidal shear with amplitude $\gamma$, in
coordinates such that the shear is aligned with the horizontal axis
($\gamma>0$) or the vertical axis ($\gamma<0$).

In order to make analytic progress with this model, we assume that
the shape function $F(\phi)$ is an even function, i.e.,
$F(\phi)=F(-\phi)$.  In other words, we assume that the galaxy is
symmetric about the horizontal axis. The shear we consider therefore
does not have an arbitrary orientation, but is either aligned with
or orthogonal to the galaxy's symmetry axis. Although not completely
general, these two cases should bound the interesting range of
shears.

Using polar coordinates in the image plane and Cartesian
coordinates in the source plane, the lens equation has the form
\begin{eqnarray}
  y_1 &=& (1+\gamma)\,r\cos\phi - F(\phi)\cos\phi + F'(\phi)\sin\phi\,,
    \label{eq:GISOeq1} \\
  y_2 &=& (1-\gamma)\,r\sin\phi - F(\phi)\sin\phi - F'(\phi)\cos\phi\,,
    \label{eq:GISOeq2}
\end{eqnarray}
and the lensing magnification is $\mu$ given by\footnote{Note that
in the absence of shear ($\gamma=0$), the magnification is simply
$\mu=(1-2\kappa)^{-1}$ and the critical curve is the isodensity
contour $\kappa=1/2$ of the galaxy (Witt et al.\ 2000; Evans \& Witt
2001, 2002; Zhao \& Pronk 2001).}
\begin{equation}
  \mu^{-1} = 1 - \gamma^2 - 2(1+\gamma\cos2\phi)\,\kappa(r,\phi)\,.
    \label{eq:GISOminv}
\end{equation}
The critical curve can be written in parametric form as
\begin{equation}
  r_c(\phi) = \frac{1+\gamma\cos2\phi}{1-\gamma^2}\ G(\phi)\,,
\end{equation}
which can be used in the lens equation to obtain a parametric
expression for the caustic.  The condition $F(\phi)=F(-\phi)$
ensures that there is always a cusp at $\phi=0$, whose location
in the image and source planes is
\begin{eqnarray}
  x_{1c} &=& \frac{G(0)}{1-\gamma}\ , \\
  y_{1c} &=& \frac{2\gamma F(0) + (1+\gamma) F''(0)}{1-\gamma}\ .
\end{eqnarray}
In general this cusp is a simple cusp, but for some combinations
of the shape function $G(\theta)$ and shear $\gamma$ it can be
part of a higher-order butterfly catastrophe.  We find that
butterfly catastrophes are rare on the major axis of the lens
potential, but can be relatively common on the minor axis when
the potential has significant power in high-order multipole
modes.

The Taylor series coefficients used to define the local orthogonal
coordinate system in \refapp{orthog} are as follows:
\begin{eqnarray}
  \ha &=& - \frac{\gamma}{2}\ ,
  \qquad
  \hb = 0\, ,
  \qquad
  \hc = \frac{1}{2}\ , \\
  a &=& \frac{(1-\gamma)^3}{6}\ \frac{3G(0)-G''(0)}{G(0)^3}\ ,
  \qquad
  c  =  1+\gamma\, , \\
  b &=& \frac{(1-\gamma)^2}{G(0)}\ ,
  \qquad
  \hp = \frac{6 G(0)^2}{(1-\gamma)[6\gamma G(0)-(1+\gamma)G''(0)]}\ .
\end{eqnarray}
Hence the transformation matrix ${\rm M}$ in \eq{coord-change} is
the identity matrix, so the $\bt$ and $\bu$ coordinate systems are
simply the $\bx$ and $\bby$ coordinate systems translated so the
cusp point is at the origin.

A source on the horizontal axis inside the caustic has at least one
image on the $x_1$-axis and two images off the $x_1$-axis. Because
of the reflection symmetry $F(\phi)=F(-\phi)$, the two off-axis
images are identical modulo some signs. The polar radius for these
two images, labeled A and C, is found by requiring that
\eq{GISOeq2} have a non-trivial solution (i.e., $\phi \ne 0$):
\begin{equation}
  r_A = r_C = \frac{1}{1-\gamma} \left[F(\phi_A)+F'(\phi_A)\cot\phi_A\right] .
\end{equation}
Their polar angles satisfy $\phi_A = -\phi_C = \theta/2$ where
$\theta$ is the opening angle defined in \refsec{configs}. The
image separation for this triplet is simply
$d = 2 r_A \sin\phi_A$.  The source position is, from \eq{GISOeq1},
\begin{equation}
  y_1 = \frac{1}{(1-\gamma)\sin\phi_A} \left[ \gamma F(\phi_A)\sin2\phi_A
    + (1+\gamma\cos2\phi_A)F'(\phi_A) \right] .
\end{equation}
Finally, the position of the on-axis image (labeled B) is found by
solving \eq{GISOeq1} with $\phi_B = 0$:
\begin{equation}
  r_B = \frac{y_1+F(0)}{1+\gamma} .
\end{equation}
Using \eq{GISOminv}, we find the magnifications of the three
images to be
\begin{eqnarray}
  \mu_{A}^{-1} = \mu_{C}^{-1} &=& 1 - \gamma^2 -
    (1-\gamma)(1+\gamma\cos2\phi_A)\,
    \frac{F(\phi_A)+F''(\phi_A)}{F(\phi_A)+F'(\phi_A)\cot\phi_A}\ ,\\
  \mu_{B}^{-1} &=& 1 - \gamma^2 - (1+\gamma)^2\,\frac{F(0)+F''(0)}{F(0)+y_1}\ .
\end{eqnarray}

\begin{figure}[t]
\centerline{\epsfxsize=3.4in \epsfbox{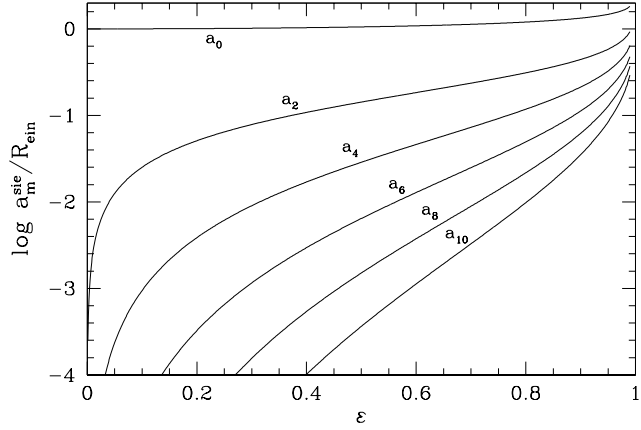}}
\caption{
The first six non-zero multipole coefficients for an SIE galaxy.
}\label{fig:sie-am}
\end{figure}

A specific case of interest is a singular isothermal ellipsoid
(SIE), which has the shape functions (Kassiola \& Kovner 1993;
Kormann, Schneider, \& Bartelmann 1994; Keeton \& Kochanek 1998)
\begin{eqnarray}
  G_{\rm sie}(\phi) &=& \frac{\Rein}{\sqrt{1-\ep\cos2\phi}}\ , \\
  F_{\rm sie}(\phi) &=& \frac{\Rein}{\sqrt{2\ep}} \left[
      \cos\phi\,\tan^{-1} \left(\frac{\sqrt{2\ep}\,\cos\phi}{\sqrt{1-\ep\cos2\phi}}\right)
    + \sin\phi\,\tanh^{-1}\left(\frac{\sqrt{2\ep}\,\sin\phi}{\sqrt{1-\ep\cos2\phi}}\right)
    \right] .
\end{eqnarray}
(The parameter $\ep$ is related to the minor-to-major axis ratio
$q$ of the ellipse by $\ep=(1-q^2)/(1+q^2)$, and it is a convenient
parameter for this formalism; however, in the main body of the paper
we always quote the true ellipticity $e=1-q$.)  When $\ep>0$ ($\ep<0$)
this formalism describes a source on the major (minor) axis of the
galaxy.  When $\ep$ and $\gamma$ have identical (opposite) signs,
the shear is aligned with (orthogonal to) the galaxy's major axis.
We note that when thinking in terms of a multipole expansion, the
SIE has power in all even multipole moments; the multipole
coefficients,
\begin{equation}
  a_m^{\rm sie} \equiv \frac{\Rein}{2\pi} \int_{0}^{2\pi}
    \frac{\cos(m\phi)}{\sqrt{1-\ep\cos2\phi}}\ d\phi\,,
\end{equation}
are shown in \reffig{sie-am}.

We also consider adding perturbations that represent departures from
elliptical symmetry in the density distribution.  Examples of such
perturbations are ``boxy'' or ``disky'' isophotes, or disk-like
components, all of which are observed (Bender et al.\ 1989; Saglia
et al.\ 1993; Kelson et al.\ 2000; Rest et al.\ 2001; Tran et al.\
2003) and predicted (Heyl et al.\ 1994; Naab \& Burkert 2003; Burkert
\& Naab 2003) in early-type galaxies.  It is convenient to express
the perturbations in terms of multipole modes.  The shape functions
for an $m$-th order mode are
\begin{eqnarray}
  G_m(\phi) &=& a_m^{\rm pert}\,\cos(m\phi)\,, \\
  F_m(\phi) &=& \frac{a_m^{\rm pert}}{1-m^2}\,\cos(m\phi)\,,
\end{eqnarray}
where the perturbation amplitude $a_m^{\rm pert}$ is defined such
that the deviation of an isodensity contour (say, $\kappa=1/2$
although since the potential is scale-free the choice is irrelevant)
from a pure ellipse is
\begin{equation}
  \delta r = a_m^{\rm pert}\,\cos(m\phi)\,.
\end{equation}
This definition is exactly equivalent to the amplitude used to
quantify isophotes shapes in observed galaxies (e.g., Bender et al.\
1989; Saglia et al.\ 1993; Rest et al.\ 2001).  It is closely related
to the fractional change in the surface density,
\begin{equation}
  \frac{\delta\kappa}{\kappa} = \frac{a_m^{\rm pert}}{\Rein}\ \cos(m\phi)
    \ \sqrt{1-\ep\cos2\phi}\ .
\end{equation}
The perturbation amplitude has dimensions of length, but we follow
observational studies and normalize it by the size of the reference
ellipse.  We can therefore adopt $a_4$ perturbation amplitudes
directly from the observational studies.

\vspace{1.2cm}
\LongTables
\input tab1

\end{document}

%% file: tab1.tex
\begin{deluxetable}{ccccrccc}
\tablecaption{Observed 4-Image Lenses}
\tablehead{
  \colhead{Lens and} &&&&&& \multicolumn{2}{c}{$\Rcusp$}
\\
  \colhead{References} &
  \colhead{Type} &
  \colhead{$\Rein\ (\arcsec)$} &
  \colhead{Triplet} &
  \colhead{$\theta\ (\arcdeg)$} &
  \colhead{$d\ (\arcsec)$} &
  \colhead{Data} &
  \colhead{Model}
}
\startdata
B0128+437
     & radio,   & 0.20 & BCD & 236.7 & 0.50 & 0.30$\pm$0.06 & 0.40 \\
(13) & fold     &      & ACD & 197.4 & 0.55 & 0.52$\pm$0.05 & 0.44 \\
     &          &      & ABD & 123.3 & 0.34 & 0.01$\pm$0.06 & 0.03 \\
     &          &      & ABC & 162.6 & 0.55 & 0.51$\pm$0.05 & 0.49 \\
\tableline
HE~0230$-$2130
    & optical, & 0.83 & A$_2$BC     & 197.2 & 1.66 & 0.35$\pm$0.06 & 0.09 \\
(1) & fold     &      & A$_1$BC     & 231.6 & 2.19 & 0.84$\pm$0.02 & 0.77 \\
    &          &      & A$_1$A$_2$C & 162.8 & 1.66 & 0.03$\pm$0.07 & 0.01 \\
    &          &      & A$_1$A$_2$B & 128.4 & 2.19 & 0.28$\pm$0.06 & 0.41 \\
\tableline
MG~0414+0534
    & near-IR, & 1.08 & A$_2$BC     & 216.0 & 2.13 & 0.40$\pm$0.05 & 0.57 \\
(1) & fold     &      & A$_1$BC     & 258.5 & 2.13 & 0.77$\pm$0.03 & 0.78 \\
    &          &      & A$_1$A$_2$C & 144.0 & 2.08 & 0.04$\pm$0.06 & 0.09 \\
    &          &      & A$_1$A$_2$B & 101.5 & 2.03 & 0.31$\pm$0.06 & 0.12 \\
\tableline
HE~0435$-$1223
     & optical, & 1.18 & BCD & 179.0 & 2.25 & 0.34$\pm$0.05 & 0.30 \\
(16) & cross    &      & ACD & 201.7 & 2.56 & 0.48$\pm$0.05 & 0.50 \\
     &          &      & ABD & 181.0 & 2.25 & 0.08$\pm$0.06 & 0.20 \\
     &          &      & ABC & 158.3 & 2.56 & 0.43$\pm$0.05 & 0.34 \\
\tableline
B0712+472
       & radio,   & 0.65 & BCD  & 200.4 & 1.27 & 0.33$\pm$0.06 & 0.62 \\
(8, 9) & cusp     &      & ACD  & 283.1 & 1.25 & 0.89$\pm$0.01 & 0.91 \\
       &          &      & ABD  & 159.6 & 1.27 & 0.06$\pm$0.07 & 0.07 \\
       &          &      & ABC* &  76.9 & 1.05 & 0.26$\pm$0.06 & 0.08 \\
\tableline
B0712+472
       & opt/IR,  & 0.65 & BCD  & 200.4 & 1.27 & 0.44$\pm$0.05 & 0.62 \\
(1, 9) & cusp     &      & ACD  & 283.1 & 1.25 & 0.69$\pm$0.03 & 0.91 \\
       &          &      & ABD  & 159.6 & 1.27 & 0.32$\pm$0.06 & 0.07 \\
       &          &      & ABC* &  76.9 & 1.05 & 0.60$\pm$0.04 & 0.08 \\
\tableline
RX~J0911+0551
    & near-IR, & 0.96 & A$_2$A$_3$B      & 179.4 & 3.26 & 0.45$\pm$0.05 & 0.44 \\
(1) & cusp     &      & A$_1$A$_3$B      & 290.4 & 3.08 & 0.59$\pm$0.04 & 0.67 \\
    &          &      & A$_1$A$_2$B      & 180.6 & 3.26 & 0.14$\pm$0.06 & 0.38 \\
    &          &      & A$_1$A$_2$A$_3$* &  69.6 & 0.96 & 0.23$\pm$0.06 & 0.00 \\
\tableline
SDSS~J0924+0219
    & optical, & 0.87 & BCD & 217.7 & 1.61 & 0.04$\pm$0.08 & 0.46 \\
(7) & cross    &      & ACD & 142.3 & 1.61 & 0.36$\pm$0.07 & 0.09 \\
    &          &      & ABD & 156.4 & 1.79 & 0.89$\pm$0.02 & 0.29 \\
    &          &      & ABC & 203.6 & 1.79 & 0.58$\pm$0.05 & 0.56 \\
\tableline
PG~1115+080
    & optical, & 1.14 & A$_2$BC     & 233.3 & 2.16 & 0.52$\pm$0.05 & 0.59 \\
(6) & fold     &      & A$_1$BC     & 218.8 & 2.43 & 0.76$\pm$0.03 & 0.70 \\
    &          &      & A$_1$A$_2$C & 141.2 & 2.43 & 0.32$\pm$0.06 & 0.18 \\
    &          &      & A$_1$A$_2$B & 126.7 & 1.86 & 0.10$\pm$0.06 & 0.06 \\
\tableline
1RXS~J1131$-$1231
     & optical, & 1.81 & BCD  & 290.5 & 3.18 & 0.87$\pm$0.01 & 0.90 \\
(15) & cusp     &      & ACD  & 181.0 & 3.20 & 0.34$\pm$0.06 & 0.31 \\
     &          &      & ABD  & 179.0 & 3.20 & 0.15$\pm$0.06 & 0.29 \\
     &          &      & ABC* &  69.5 & 2.38 & 0.35$\pm$0.06 & 0.06 \\
\tableline
HST~12531$-$2914
        & optical, & 0.54 & BCD & 172.4 & 1.04 & 0.39$\pm$0.05 & 0.28 \\
(1, 14) & cross    &      & ACD & 187.6 & 1.04 & 0.46$\pm$0.05 & 0.14 \\
        &          &      & ABD & 206.8 & 1.36 & 0.31$\pm$0.06 & 0.59 \\
        &          &      & ABC & 153.2 & 1.36 & 0.17$\pm$0.06 & 0.33 \\
\tableline
HST~14113+5211
    & optical, & 0.83 & BCD & 168.9 & 1.35 & 0.14$\pm$0.06 & 0.09 \\
(4) & cross    &      & ACD & 161.3 & 2.28 & 0.36$\pm$0.05 & 0.42 \\
    &          &      & ABD & 191.1 & 1.42 & 0.15$\pm$0.06 & 0.16 \\
    &          &      & ABC & 198.7 & 2.28 & 0.67$\pm$0.03 & 0.63 \\
\tableline
H1413+117
    & near-IR, & 0.56 & BCD & 198.6 & 1.35 & 0.48$\pm$0.05 & 0.28 \\
(1) & cross    &      & ACD & 186.2 & 1.10 & 0.37$\pm$0.05 & 0.83 \\
    &          &      & ABD & 173.8 & 1.10 & 0.26$\pm$0.06 & 0.78 \\
    &          &      & ABC & 161.4 & 1.35 & 0.24$\pm$0.06 & 0.61 \\
\tableline
HST~14176+5226
        & optical, & 1.33 & BCD & 172.8 & 2.36 & 0.04$\pm$0.06 & 0.03 \\
(1, 14) & cross    &      & ACD & 198.1 & 3.26 & 0.54$\pm$0.04 & 0.64 \\
        &          &      & ABD & 187.2 & 2.36 & 0.13$\pm$0.06 & 0.08 \\
        &          &      & ABC & 161.9 & 3.26 & 0.54$\pm$0.04 & 0.50 \\
\tableline
B1422+231
        & radio,   & 0.76 & BCD  & 187.2 & 1.29 & 0.35$\pm$0.06 & 0.35 \\
(5, 12) & cusp     &      & ACD  & 283.0 & 1.29 & 0.96$\pm$0.01 & 0.94 \\
        &          &      & ABD  & 172.8 & 1.25 & 0.05$\pm$0.07 & 0.15 \\
        &          &      & ABC* &  77.0 & 1.29 & 0.18$\pm$0.06 & 0.12 \\
\tableline
B1555+375
     & radio,   & 0.23 & BCD & 209.3 & 0.42 & 0.14$\pm$0.07 & 0.56 \\
(11) & fold     &      & ACD & 257.4 & 0.42 & 0.90$\pm$0.01 & 0.89 \\
     &          &      & ABD & 150.7 & 0.42 & 0.21$\pm$0.06 & 0.03 \\
     &          &      & ABC & 102.6 & 0.41 & 0.45$\pm$0.05 & 0.14 \\
\tableline
B1608+656
     & radio,   & 0.72 & BCD & 191.5 & 2.04 & 0.16$\pm$0.06 & 0.06 \\
(10) & fold     &      & ACD & 168.5 & 2.04 & 0.19$\pm$0.06 & 0.24 \\
     &          &      & ABD & 261.0 & 2.10 & 0.79$\pm$0.02 & 0.89 \\
     &          &      & ABC &  99.0 & 2.10 & 0.49$\pm$0.05 & 0.49 \\
\tableline
B1933+503
    & radio,   & 0.49 & 3,4,6 & 143.0 & 0.82 & 0.39$\pm$0.05 & 0.01 \\
(2) & fold     &      & 1,4,6 & 199.7 & 1.16 & 0.70$\pm$0.03 & 0.63 \\
    &          &      & 1,3,6 & 217.0 & 0.91 & 0.21$\pm$0.06 & 0.29 \\
    &          &      & 1,3,4 & 160.3 & 1.16 & 0.72$\pm$0.03 & 0.42 \\
\tableline
B2045+265
    & radio,   & 1.13 & BCD  & 183.9 & 1.93 & 0.05$\pm$0.06 & 0.49 \\
(3) & cusp     &      & ACD  & 325.1 & 1.93 & 0.88$\pm$0.01 & 0.98 \\
    &          &      & ABD  & 176.1 & 1.92 & 0.21$\pm$0.06 & 0.19 \\
    &          &      & ACD* &  34.9 & 0.84 & 0.52$\pm$0.04 & 0.02 \\
\tableline
Q2237+030
    & optical, & 0.85 & BCD & 186.5 & 1.65 & 0.29$\pm$0.06 & 0.18 \\
(1) & cross    &      & ACD & 173.5 & 1.65 & 0.20$\pm$0.06 & 0.12 \\
    &          &      & ABD & 146.2 & 1.83 & 0.71$\pm$0.03 & 0.39 \\
    &          &      & ABC & 213.8 & 1.83 & 0.52$\pm$0.05 & 0.62
\enddata
\tablecomments{
Results for image triplets in the nineteen published 4-image lenses.
Column 2 gives the image configuration (fold, cusp, or cross) and
indicates whether the flux ratios are measured at optical, near-IR,
or radio wavelengths.  The uncertainties in the observed values of
$\Rcusp$ are obtained by assuming 10\% uncertainties in the image
fluxes; see \refsec{obs} for more discussion.  The predicted values
of $\Rcusp$ are computed with standard lens models.  For the cusp
lenses B0712+472, RX~J0911+0551, 1RXS~J1131$-$1231, B1422+231, and
B2045+265, the cusp image triplet is indicated by *.  Note that
B0712+472 appears twice because we report data from both radio and
optical/near-IR wavelengths.  The references are as follows:
(1) CASTLES (see {\tt http://cfa-www.harvard.edu/castles});
(2) Cohn et al.\ 2001;
(3) Fassnacht et al.\ 1999;
(4) Fischer et al.\ 1998;
(5) Impey et al.\ 1996;
(6) Impey et al.\ 1998;
(7) Inada et al.\ 2003;
(8) Jackson et al.\ 1998;
(9) Jackson et al.\ 2000;
(10) Koopmans \& Fassnacht 1999;
(11) Marlow et al.\ 1999;
(12) Patnaik et al.\ 1999;
(13) Phillips et al.\ 2000;
(14) Ratnatunga et al.\ 1995;
(15) Sluse et al.\ 2003;
(16) Wistoski et al.\ 2002.
}\label{tab:data}
\end{deluxetable}